\documentclass[aps,prc,floatfix,twocolumn,superscriptaddress,showpacs]{revtex4-1}

\usepackage{graphicx}
\usepackage{dcolumn}
\usepackage{amsmath}
\usepackage{amssymb}
\usepackage{bm}
\usepackage[colorlinks,citecolor=blue]{hyperref}

\newcommand{\matrixelementreduceds}[3]{\left\langle #1 \left\lVert #2 \right\rVert #3 \right\rangle}

\begin{document}

\title{Large scale evaluation of beta-decay rates of r-process
  nuclei with the inclusion of first-forbidden transitions}

\author{T. Marketin}
\affiliation{Physics Department, Faculty of Science, University of
  Zagreb, 10000 Zagreb, Croatia} 
\author{L. Huther}
\affiliation{Institut f\"{u}r Kernphysik (Theoriezentrum), Technische Universit\"{a}t Darmstadt, 64289 Darmstadt, Germany}
\author{G. Mart\'{i}nez-Pinedo}
\affiliation{Institut f\"{u}r Kernphysik (Theoriezentrum), Technische Universit\"{a}t Darmstadt, 64289 Darmstadt, Germany}
\affiliation{GSI Helmholtzzentrum f\"{u}r Schwerioneneforschung, Planckstra{\ss}e 1, 64291 Darmstadt, Germany
}

\date{\today}

\begin{abstract}
\begin{description}
\item[Background] R-process nucleosynthesis models rely, by necessity,
  on nuclear structure models for input. Particularly important are
  beta-decay half-lives of neutron rich nuclei. At present only a single
  systematic calculation exists that provides values for all
  relevant nuclei making it difficult to test the sensitivity of
  nucleosynthesis models to this input. Additionally, even though
  there are indications that their contribution may be significant,
  the impact of first-forbidden transitions on decay rates has not
  been systematically studied within a consistent model.

\item[Purpose] To provide a table of $\beta$-decay half-lives and
  $\beta$-delayed neutron emission probabilities, including
  first-forbidden transitions, calculated within a fully
  self-consistent microscopic theoretical framework. The results
  are used in an r-process nucleosynthesis calculation to asses
  the sensitivity of heavy element nucleosynthesis to weak interaction
  reaction rates.

\item[Method] We use a fully self-consistent covariant density
  functional theory (CDFT) framework. The ground state of all nuclei
  is calculated with the relativistic Hartree-Bogoliubov (RHB) model,
  and excited states are obtained within the proton-neutron
  relativistic quasiparticle phase approximation (pn-RQRPA). 

\item[Results] The $\beta$-decay half-lives, $\beta$-delayed neutron
  emission probabilities and the average number of emitted neutrons
  have been calculated for 5409 nuclei in the neutron-rich region of
  the nuclear chart. We observe a significant contribution of the
  first-forbidden transitions to the total decay rate in nuclei far from the valley of stability. The experimental half-lives are in
  general well reproduced, both for even-even, odd-A and odd-odd
  nuclei, in particular for short-lived nuclei. 

\item[Conclusions] In certain regions of the nuclear chart,
  first-forbidden transitions constitute a large fraction of the total
  decay rate and must be taken into account consistently in modern
  evaluations of half-lives. Both the beta-decay half-lives, and beta-delayed neutron emission probabilities have a noticeable
  impact on the results of heavy element nucleosynthesis models.

\end{description}
\end{abstract}


\maketitle
\section{Introduction}

One of the currently most active fields of nuclear astrophysics deals
with the synthesis of elements heavier than iron by the
r-process. Even though its astrophysical site has not yet been
identified, it is commonly accepted that it occurs in explosive
environments involving relatively high temperatures (up to 1~GK) and
neutron densities ($n_n > 10^{20}$~g~cm$^{-3}$). Apart from the
complexities involved in the astrophysical modeling, the
r-process represents a particularly difficult challenge due to the
large amount of nuclear input required. It is a complex, dynamical
process involving a delicate interplay between the strong,
electromagnetic and weak interactions, requiring the knowledge of a
number of observables in several thousand nuclei across the whole
nuclear chart. One of the crucial nuclear properties having a direct
impact on the distribution of elemental abundances is the
$\beta$-decay half-lives of the participating
nuclei~\cite{Langanke2003,Arnould2007}. It determines the speed of
matter flow towards higher atomic numbers, setting the time scale for
the r-process. $\beta$-delayed neutron emission occurs during the
whole r-process duration but becomes particularly important at late
phases when a competition between neutron captures and beta-decays
takes place~\cite{Arcones2011}. Particularly important are the
beta-decay half-lives of nuclei around magic neutron numbers $N=50,
82$ and 126. The neutron separation energies show discontinuities
around these magic numbers resulting in rather low neutron capture
rates. As a consequence the r-process matter flow moves closer to the
stability where nuclei have substantially larger beta-decay
half-lives. Thus matter accumulates around magic numbers $N=50, 82$
and 126 producing the observed peaks in the solar system r-process
distribution.

At present only a few half-lives of r-process nuclei in the vicinity
of the magic numbers
$N=50$~\cite{Hosmer2005,Hosmer2010,Quinn2012,Madurga2012} and
$N=82$~\cite{Montes2006,Nishimura2011,Dillmann2003,Pfeiffer2001} have
been measured. Even though the experiments at GSI have provided
valuable information for half-lives approaching the third r-process
peak~\cite{Benlliure2012,Benzoni2012,Domingo-Pardo.Caballero-Folch.ea:2013,Kurtukian-Nieto.Benlliure.ea:2014,Morales.Benlliure.ea:2014},
so far no experimental half-lives for r-process nuclei at the $N=126$
shell closure are available. Hence, r-process nucleosynthesis
calculations rely mainly on theoretical half-lives. The interacting
shell model~\cite{Caurier2005} has recently been extended to include
first-forbidden transitions and has been applied to the calculation of
half-lives for r-process nuclei around $N=50$, 82 and 126 shell
closures~\cite{Zhi2013}. However, due to the increasing computational
cost, shell-model calculations are restricted to nuclei near closed
shells. For this reason most of the half-lives calculations for
r-process nuclei are based on the quasiparticle random phase
approximation on top of semiempirical global
models~\cite{Moeller2003,Borzov2006,Fang2013a}, the
Hartree-Fock-Bogoliubov~\cite{Engel1999}, or the relativistic
Hartree-Bogoliubov~\cite{Niksic2005,Marketin2007,Niu2013a}. The
quasiparticle random phase approximation (QRPA) approach based on a
global effective interaction presents a valid alternative to the
interacting shell model. It has been successfully employed in numerous
applications~\cite{Paar2007} and can provide a systematic description
of $\beta$-decay properties of arbitrarily heavy nuclei. The
importance of performing calculations based on the self-consistent
mean-field models, as opposed to the empirical mean-field potentials
has been already emphasized~\cite{Borzov2006}. Currently, however, the
only available large-scale calculation of $\beta$-decay properties of
nuclei used in heavy element nucleosynthesis simulations is based on
an QRPA calculation based on a schematic interaction on top of the
Finite Range Droplet Model~\cite{Moeller2003}. Additionally, in
$\beta$-decay calculations the impact of first-forbidden transitions
on the total decay rate is mostly unexplored. While these transitions
and their higher order counterparts have been included in the
investigations of electron and muon capture rates and neutrino-nucleus interactions, their use in $\beta$-decay studies has been  either limited to particular isotopic chains~\cite{Borzov2003a} or have been described using a different theoretical foundation such as
the gross theory~\cite{Moeller2003}. Furthermore, due to the existence
of a single data table of $\beta$-decay half-lives, it is not possible
to perform meaningful sensitivity studies on the influence of
beta-decay rates on r-process nucleosynthesis. 

In an effort to avoid effects coming from combining models built on
differing theoretical bases, and with the aim of providing a new, high
precision data table of $\beta$-decay properties, in this study we
utilize a fully self-consistent theoretical framework based on the
relativistic Hartree-Bogoliubov (RHB) model~\cite{Vretenar2005a} for
the description of the ground state of open- and closed-shell nuclei
with the proton-neutron relativistic QRPA (pn-RQRPA) where the
residual interaction is derived from the same density functional as
was used for the ground state calculations. In this way, we ensure the
consistency among the various required properties, from the masses and
$Q_{\beta}$ values to the nuclear response strength functions. The RHB
+ pn-RQRPA was already successfully employed in the study of
Gamow-Teller and higher order resonances~\cite{Paar2004,
  Marketin2012a}, $\beta$-decay
half-lives~\cite{Niksic2005,Marketin2007} and stellar weak-interaction
processes~\cite{Paar2008,Niu2011}. This framework also enables the
treatment of first-forbidden transitions on an equal footing with the
Gamow-Teller transitions, allowing for a meaningful investigation of
the contribution of the $J > 0$ modes of nuclear excitation and for
studying the regions of interest where this contribution is
significant. We use this model to calculate the $\beta$-decay
half-lives and the $\beta$-delayed neutron emission probabilities of
5409 neutron-rich nuclei with $8 \le Z \le 124$, including even-nuclei
but also nuclei with an odd number of particles. In
Section~\ref{sec:theory} we present the theoretical formalism of the
model, while Section~\ref{sec:results} contains all the necessary
expressions for the evaluation of matrix elements and nuclear
properties, together with the presentation and discussion of obtained
results. Finally, in Section~\ref{sec:conclusion} we provide the
concluding remarks.

\section{Theoretical formalism} \label{sec:theory}

\subsection{QRPA calculations}
\label{sec:qrpa-calculations}

The calculation of beta-decay rates requires the calculation of both
the nuclear ground state and excited states of the daughter nucleus
and the transitions between them, together with the evaluation of the
lepton phase space involved in the transition. Because the calculation
requires a good description of a wide range of physical quantities,
and because the goal is to obtain the decay rates for a very large
range of nuclei, we employ a fully microscopic theoretical framework
based on the relativistic nuclear energy density functional
(RNEDF). The RNEDF based framework employs the self-consistent mean
field for nucleons and a minimal set of meson fields; the isoscalar
scalar $\sigma$ meson ($J^{\pi}= 0^{+} , T = 0$), the isoscalar vector
$\omega$ meson ($J^{\pi} = 1^{-} , T = 0$) and the isovector vector
$\rho$ meson ($J^{\pi} = 1^{-} , T = 1$), supplemented with the
electromagnetic field. The meson-nucleon interaction is included with
a minimal set of the interaction terms, where the vertex functionals
include explicit dependence on the nucleon vector
density~\cite{Vretenar2005a}. The nuclear ground state properties are
described using the relativistic Hartree-Bogoliubov model (RHB), which
properly describes the pairing effects in open-shell nuclei. For the
model parameters that determine the density-dependent couplings and
the meson masses, in this work the D3C$^{*}$ parametrization is
used~\cite{Marketin2007}, as it was previously shown that it provides
a good description of $\beta$-decay half-lives in medium and heavy
nuclei. The pairing correlations in open shell nuclei are described by
the finite range Gogny interaction, with the D1S
parametrization~\cite{Berger1984}.

The excited states are obtained using the proton-neutron relativistic
quasiparticle random phase approximation (pn-RQRPA), formulated in the
canonical single-nucleon basis of the RHB model~\cite{Paar2003} and
extended to the description of charge-exchange excitations
(pn-RQRPA)~\cite{Paar2004}. The RHB + RQRPA model is fully
self-consistent in both the \emph{ph} and \emph{pp} channels. The same
interactions are used in the RHB equations that determine the
canonical quasiparticle basis, and in the matrix equations of the
RQRPA. Transitions between the $0^{+}$ ground state of a spherical
parent nucleus and the $J^\pi$ excited state of the corresponding
daughter nucleus are induced by a charge-exchange operator
$T^{JM}$. Assuming spherical symmetry of the nuclear system, the
quasiparticle pairs can be coupled to good angular momentum and the
matrix equations of the pn-RQRPA read~\cite{Ring1980}:

\begin{equation} \left( \begin{array} [c]{cc}
A^{J} & B^{J}\\
B^{^{\ast}J} & A^{^{\ast}J}
\end{array}
\right)  \left( \begin{array} [c]{c}
X^{\lambda J}\\
Y^{\lambda J}
\end{array}
\right)  =E_{\lambda}\left( \begin{array} [c]{cc}
1 & 0\\
0 & -1
\end{array}
\right)  \left( \begin{array} [c]{c}
X^{\lambda J}\\
Y^{\lambda J}
\end{array}\right) \; , 
\label{eq:pnrqrpaeq}
\end{equation}
where the matrices $A$ and $B$ are defined in the canonical
basis~\cite{Ring1980}. For each energy $E_{\lambda}$, $X^{\lambda J}$
and $Y^{\lambda J}$ in Eq. (\ref{eq:pnrqrpaeq}) denote the
corresponding forward- and backward-going QRPA amplitudes,
respectively. The transition matrix element between the ground
state of the parent nucleus and the excited state of daughter nucleus,
induced by the operator $T^{JM}$, reads 
\begin{equation} \label{eq:strength-}
B_{\lambda J}^{\pm} = \left| \sum_{pn} \langle p||T^J||n \rangle \left(
    X_{pn}^{\lambda J} u_p v_n + (-1)^J Y_{pn}^{\lambda J}v_p u_n
  \right) \right|^2 \; . 
\end{equation}
where the $X$ and $Y$ amplitudes are obtained from solving the
pn-RQPA equation~\eqref{eq:pnrqrpaeq}.

In the $T = 1$ channel of the pn-RQRPA we use the pairing part of the
Gogny force:
\begin{eqnarray}
\label{eq:gogny}
V^{pp}(1,2) = \sum_{i=1,2}e^{-r_{12}^2/\mu_{i}^{2}}\,(W_{i}&+&B_{i}P^{\sigma }-H_{i}P^{\tau}\nonumber\\
& & -M_{i}P^{\sigma }P^{\tau }), 
\end{eqnarray}
with $r_{12} = |\bm{r}_1 - \bm{r}_2|$, $P^\sigma, P^\tau$ the spin and
isospin exchange operators and the D1S set of parameters $\mu _{i}$,
$W_{i}$, $B_{i} $, $H_{i}$ and $M_{i}$ $(i=1,2)$ taken
from~\cite{Berger1984} . This force has been carefully adjusted to pairing properties of finite nuclei all over the periodic table. In
particular, the basic advantage of the Gogny force is the finite
range, which automatically guarantees a proper cut-off in the momentum
space. For the $T = 0$ proton-neutron pairing interaction in open
shell nuclei we use a form consisting of a short-range repulsive
Gaussian combined with a weaker longer-range attractive Gaussian.
\begin{equation} \label{eq:residual}
V_{12}=-V_{0}\sum_{j=1}^{2}
g_{j}e^{-\frac{r_{12}^{2}}{\mu_{j}^{2}}}\hat{\Pi}_{S=1,T=0}. 
\end{equation}
where $\hat\Pi_{S=1,T=0}$ projects onto states with $S=1$ and
$T=0$. The ranges $\mu_{1}=1.2$~fm and $\mu_{2}=0.7$~fm of the two
Gaussians are taken from the Gogny interaction. The relative strengths
$g_{1} =1$ and $g_{2} = -2$ are chosen so that the force is repulsive
at small distances. The only remaining free parameter is $V_{0}$, the
overall strength. This interaction, with a constant value of $V_{0}$
was used in the non-relativistic QRPA calculation \cite{Engel1999} of
$\beta$-decay rates for spherical neutron-rich $r$-process
waiting-point nuclei. Two relativistic calculations of $\beta$-decay
half-lives of neutron-rich nuclei~\cite{Niksic2005,Marketin2007} have
shown that a single value of the overall interaction strength cannot
be successfully used in different regions of the nuclear chart. Thus,
we take the ansatz proposed in~\cite{Niu2013a}
\begin{equation} \label{eq:t0pairing}
V_{0} = V_{L} + \frac{V_{D}}{1 + e^{a + b\left(N-Z\right)}},
\end{equation}
with values $V_{L} = 160.0$~MeV, $V_{D} = 15.0$ MeV, $a = 7.2$ and $b
= -0.3$ adjusted to obtain the best possible description of available
half-life data~\cite{Audi2012}.  

\begin{figure}[htb]
  \centering
  \includegraphics[width=\linewidth]{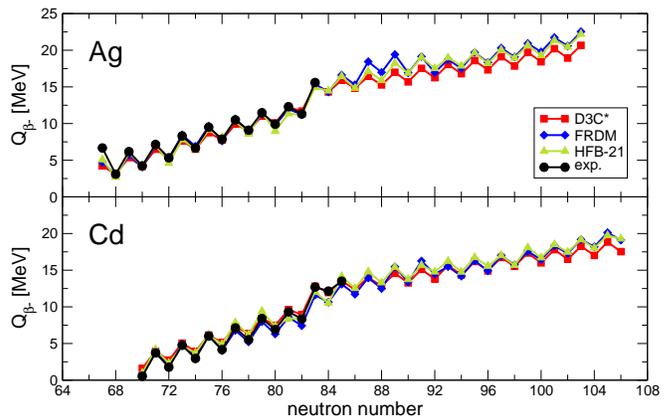}
  \caption{\label{fig:Qval} (color online) $\beta$-decay Q-values for
    the silver and cadmium isotopic chains. Values calculated in this
    work, marked with red squares, are compared with the values
    obtained from the FRDM mass table~\cite{Moeller1995}, denoted with
    blue diamonds, the HFB-21 model~\cite{Goriely2010} and with existing experimental data~\cite{Audi2012} denoted with black circles. }
\end{figure}

The model employed in this study is well suited to describe the
properties of the ground state and of the excited states of even-even
nuclei. However, as the r-process naturally includes odd-A and odd-odd
nuclei as well, it becomes necessary to provide at least an
approximate description of the decay properties of these nuclei. With
this aim, we compute the ground state of odd nuclei by employing the
same model as for even-even nuclei, but constraining the expectation
value of the particle number operator to an odd number of protons
and/or neutrons. In this way an \emph{even} RHB state is obtained,
whose energy is different from \emph{true} odd nucleus ground state
energy by energy of the odd
quasi-particle~\cite{Duguet2001,Duguet2001a}.  To quantify the
validity of this approximation we show the computed $Q_{\beta}$ values
for the silver and cadmium ($Z=45$) isotopic chains in
Fig.~\ref{fig:Qval}, calculated by approximating the binding energy of
the daughter nucleus~\cite{Engel1999}:

\begin{eqnarray} \label{eq:qval}
Q & = & \Delta M_{nH} + \lambda_{n} - \lambda_{p} - E_{2qp}  \nonumber \\
  & = & M_{ex}(Z,N) - M_{ex}(Z+1,N-1), 
\end{eqnarray}
where $\lambda_{n}$ and $\lambda_{p}$ are the neutron and proton Fermi
energies in the ground state of the parent nucleus, $E_{2qp}$ are the
lowest quasiparticle energies. $\Delta M_{nH} = 0.782$~MeV is the mass
difference between a neutron and the hydrogen atom, and $M_{ex}(Z,N)$
is the mass excess. In silver isotopes (odd Z), the $Q_{\beta}$ values
are well reproduced for nuclei with an even number of neutrons, and
only slightly underestimated for odd-odd nuclei. The predicted values
closely follow the jump above the closed neutron shell, and begin to
deviate from the FRDM values after the shell closure, while the HFB-21 results are found between the other two models. For the cadmium
isotopes the model overestimates the $Q_{\beta}$ values in nuclei with
an even number of neutrons in lighter isotopes, but above $N = 78$ the
results reproduce the experimental values accurately. Again, the model
very successfully reproduces the changes around the closed neutron
shell. For heavier isotopes there are no available data, but all the models coincide in their predictions of the $Q_{\beta}$ values. We can conclude from these results that this approach to estimating the Q-values provides a good agreement with the available data and does not hinder our study of weak-interaction processes.

\subsection{Beta-decay half-lives}
\label{sec:beta-decay-half}

In the present study of beta-decay half-lives we include both allowed ($L = 0$)
and first-forbidden ($L = 0, 1$) transitions. The beta-decay rate for
a transition between an initial and final nuclear state is equal
to~\cite{Behrens1969}  
\begin{equation} \label{eq:rate}
\lambda = \frac{\ln 2}{K}
\int_{0}^{p_{0}} p_{e}^{2} \left(W_{0} - W \right)^{2} F(Z,W) C(W)
dp_{e}, 
\end{equation}
where $W$ is the electron energy in units of $m_{e}c^{2}$ with $W_{0}$
being the maximum electron energy that is equal to the difference in
nuclear masses between initial and final nuclear states, $W_0 = (M_i -
M_f)/m_e$, and $p_e$ is the electron momentum in units of $m_e c$. We
approximate the maximum electron energy with~\cite{Engel1999}:
\begin{equation}
M_{i} - M_{f} \approx \lambda_{n} - \lambda_{p} + \Delta m_{np} - E_{QRPA}.
\end{equation}
The constant $K$ is measured in superallowed beta-decay to be,
$K=6144\pm 2$~s~\cite{Hardy2009}.

The shape factor $C(W)$ differs for various decays. For allowed decays
it is energy independent. In the case of $\beta^-$ decay of neutron-rich
nuclei the shape factor is simply the Gamow-Teller reduced transition
probability, $C(W)=B(GT)$, with
\begin{equation}
  \label{eq:bgt}
  B(GT) = g_A^2
  \frac{\matrixelementreduceds{f}{\sum_k \bm{\sigma}^k \bm{t}^k_- }{i}^2}{(2J_i+1)}.
\end{equation}
The matrix element is reduced with respect to the spin operator
$\bm{\sigma}$ only using the Condon-Shortley phase
convection~\cite{Edmonds:1960}. The sum runs over all nucleons. For
the isospin lowering operator we use the convention $t_-
\left|n\right\rangle= \left|p\right\rangle$. Finally,
$g_A=-1.2701(25)$~\cite{Beringer.Arguin.ea:2012} is the weak axial
coupling constant.

For first-forbidden transitions the shape factor reads:
\begin{equation}
C(W) = k + ka W + kb/W + kc W^{2}.
\end{equation}
where the factors $k$, $ka$, $kb$ and $kc$ are defined as~\cite{Behrens1971}:
\begin{subequations}
\begin{eqnarray}
k  & = & \left[\zeta_{0}^{2} + \frac{1}{9}w^{2} \right]_{(0)} +
\left[\zeta_{1}^{2} + \frac{1}{9}(x + u)^{2} - \frac{4}{9}\mu_{1}
  \gamma_{1} u(x + u) \right. \nonumber \\ 
   &   & \left. + \frac{1}{18}W_{0}^{2}(2x + u)^{2} -
     \frac{1}{18}\lambda_{2}(2x - u)^{2} \right]_{(1)} \\
   & & + \left[ \frac{1}{12} z^{2} \left(W_{0}^{2} - \lambda_{2} \right) \right]_{(2)}, \nonumber \\
ka & = & \left[-\frac{4}{3}uY - \frac{1}{9}W_{0}\left( 4x^{2} + 5u^{2}\right) \right]_{(1)} - \left[ \frac{1}{6} W_{0} z^{2} \right]_{(2)}, \\
kb & = & \frac{2}{3}\mu_{1}\gamma_{1}\left\{ -\left[\zeta_{0}w\right]_{(0)} + \left[\zeta_{1}(x + u)\right]_{(1)} \right\}, \\
kc & = & \frac{1}{18}\left[ 8u^{2} + (2x + u)^{2} + \lambda_{2}(2x -
  u)^{2} \right]_{(1)} \\
& & + \frac{1}{12}\left[\left(1 + \lambda_{2}\right)z^{2}\right]_{(2)},\nonumber
\end{eqnarray}
\end{subequations}
with
\begin{equation}
  \label{eq:coefdef}
  \begin{array}{ll}
    V=\xi^\prime v + \xi w^\prime, & \zeta_0=V+\frac{1}{3} w W_0,\\[2mm]
    Y=\xi^\prime y - \xi (u^\prime+ x^\prime), & \zeta_1= Y +
    \frac{1}{3} (u-x) W_0.
  \end{array}
\end{equation}

The numbers in parenthesis after the closing brackets denote the rank
of the operators inside. The parameter $\gamma_1$ is given by
$\sqrt{1-(\alpha Z)^2}$. For the coulomb wave functions we use the
approximations $\mu_1 \approx 1$ and $\lambda_2\approx
1$~\cite{Behrens1982}. The quantity $\xi =
\alpha Z/(2 R)$ where we choose $R =\sqrt{\langle r\rangle^2}$.

In the Condon-Shortley phase convention~\cite{Edmonds:1960} the matrix
elements for $\beta^-$ transitions are:
\begin{subequations}
\begin{eqnarray}
w & = & -g_A \sqrt{3} \frac{\matrixelementreduceds{f}{\sum_k r_k \left[
    \bm{C}^k_{1} \otimes \bm{\sigma}^k\right]^{0}
  \bm{t}^k_-}{i}}{\sqrt{2J_i+1}}, \\ 
x & = & - \frac{\matrixelementreduceds{f}{\sum_k
    r_k\bm{C}^k_{1}\bm{t}^k_-}{i}}{\sqrt{2J_i+1}}, \\ 
u & = & -g_A \sqrt{2} \frac{\matrixelementreduceds{f}{\sum_k r_k \left[
    \bm{C}^k_{1} \otimes \bm{\sigma}^k\right]^{1}
  \bm{t}^k_-}{i}}{\sqrt{2J_i+1}}, \\  
z & = &  2 g_A \frac{\matrixelementreduceds{f}{\sum_k r_k \left[ \bm{C}^k_{1}
    \otimes \bm{\sigma}^k\right]^{2} \bm{t}^k_-}{i}}{\sqrt{2J_i+1}}, \\ 
w' & = & -g_A \frac{2}{\sqrt{3}}
\frac{\matrixelementreduceds{f}{\sum_k r_k I(1,1,1,1,r_k) \left[ \bm{C}^k_{1}
    \otimes \bm{\sigma}^k\right]^{0} \bm{t}^k_-}{i}}{\sqrt{2J_i+1}}, \\ 
x' & = & - \frac{2}{3} \frac{\matrixelementreduceds{f}{\sum_k r_k I(1,1,1,1,r_k)
  \bm{C}^k_{1}\bm{t}^k_-}{i}}{\sqrt{2J_i+1}}, \\ 
u' & = & - g_A  \frac{2\sqrt{2}}{3}
\frac{\matrixelementreduceds{f}{\sum_k r_k
  I(1,1,1,1,r_k) \left[ \bm{C}^k_{1} \otimes \bm{\sigma}^k\right]^{1}
  \bm{t}^k_-}{i}}{\sqrt{2J_i+1}}.
\end{eqnarray}
\end{subequations}
The matrix elements connected with the relativistic corrections are 
\begin{subequations}
\label{eq:rela}
\begin{eqnarray}
\xi'v & = & - g_A \frac{\matrixelementreduceds{f}{\sum_k \gamma^k_{5}
    \bm{t}^k_-}{i}}{\sqrt{2J_i+1}}, \\
\xi'y & = & - \frac{\matrixelementreduceds{f}{\bm{\alpha}^k\bm{t}^k_-}{i}}{\sqrt{2J_i+1}},
\end{eqnarray}
\end{subequations}
with $\gamma_5$ and $\bm{\alpha}$
being the Dirac matrices. Due to the 
fact that our formalism is relativistic we do not perform a
non-relativistic reduction in the evaluation of the matrix elements in
equations~\eqref{eq:rela}. The quantity $\bm{C}_{LM}$ equals
\begin{equation}
\bm{C}_{LM} = \sqrt{\frac{4\pi}{2L+1}} Y_{LM},
\end{equation}
where $Y_{LM}$ are the spherical harmonics

As the matrix elements are independent of the electron and neutrino
energies, the integrals over the electron phase space are evaluated
independently and appear only as multiplicative factors. The function
$I(1,1,1,1,r)$ takes into account the nuclear charge distribution, and
in the approximation of the uniform spherical distribution it has the
form~\cite{Behrens1971} 
\begin{equation}
I(1,1,1,1,r) = \frac{3}{2} \left\{ \begin{array}{rl}
1 - \frac{1}{5}\left(\frac{r}{R}\right)^{2} & 0 \le r \le R \\
\frac{R}{r} - \frac{1}{5}\left(\frac{R}{r}\right)^{3} & r > R \end{array} \right. .
\end{equation}

Systematic calculations of beta-decay half-lives determined by
Gamow-Teller transitions have shown that the theoretical matrix
elements need to be quenched by a factor $q$ that has been found to be
independent of the particular transition considered and approximately
constant over the nuclear
chart~\cite{Wildenthal.Curtin.Brown:1993,Martinez-Pinedo.Poves.ea:1996b}. There
is also evidence that the total first-forbidden transition strength is
overestimated by theoretical approaches. In addition, the quenching
factor seems to depend on the model space used and on the particular
first-forbidden
operator~\cite{Warburton:1990,Warburton1991,Warburton.Becker.ea:1988,Warburton.Towner:1994,Warburton.Towner:1994,Zhi2013,Suzuki.Yoshida.ea:2012}. So
far no study has addressed the quenching of first-forbidden
transitions of global calculations of beta-decay half-lives. In our
calculations, we account for the quenching of both the Gamow-Teller and
first-forbidden transitions by using the same effective value of $g_A =
-1.0$.

\section{ \label{sec:results} Results}

\subsection{$\beta$-decay half-lives}

\begin{figure}[htb]
  \centering
  \includegraphics[width=\linewidth]{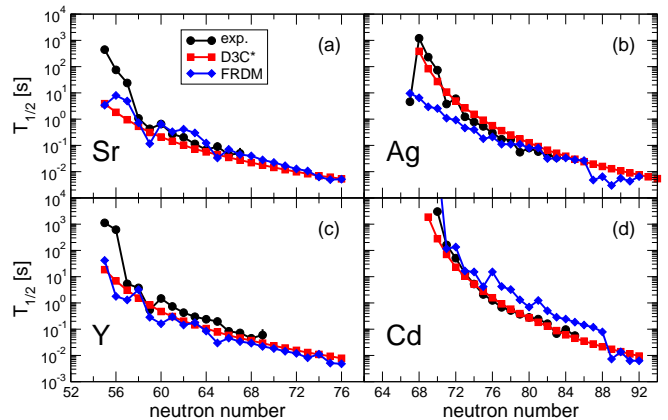}
  \caption{\label{fig:beta_chain} (color online) Comparison of the
    calculated half-lives with the FRDM results and the experimental
    data for the strontium (upper left), yttrium (lower left), silver
    (upper right) and cadmium (lower right) isotopic chains. In cases
    where they are not visible, the error bars are smaller than the
    data marker.}
\end{figure}

In Fig. \ref{fig:beta_chain} we compare the calculated half-lives for
four isotopic chains, including both the even and odd atomic numbers,
where the half-lives are obtained from the decay rate in
Eq. (\ref{eq:rate}):
\begin{equation} \label{eq:halflife}
T_{1/2} = \frac{\ln 2}{\lambda}.
\end{equation}

\begin{figure}[htb]
  \centering
  \includegraphics[width=\linewidth]{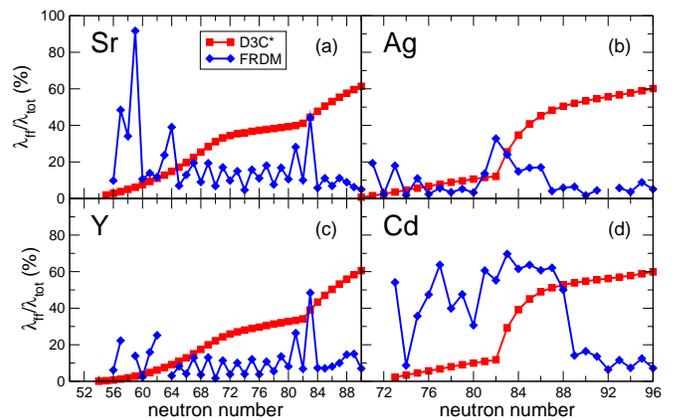}
  \caption{(color online) Percentage of first-forbidden transitions contributing to
    the decay of Sr, Y, Ag and Cd isotopes. In the cases where it was not possible to extract the value for a particular model, the data point was not plotted. \label{fig:ff_chain}} 
\end{figure}

In the four panels we show a comparison of the present results with
the data and the FRDM results for the strontium (Sr, $Z = 38$),
yttrium (Y, $Z = 39$), silver (Ag, $Z = 47$), and cadmium (Cd, $Z =
48$) isotopic chains. The two lightest isotopic chains, strontium and
yttrium, are found between two closed proton shells $Z = 28$ and $Z =
50$. Our calculation has difficulties reproducing the half-lives of
the longest lived nuclei due to the rather low Q-values, however for
isotopes more neutron-rich than $N = 58$ we obtain very good agreement
with the data, with a slight tendency to underestimate the half-lives
of yttrium isotopes. The FRDM+QRPA approach displays erratic jumps in
the half-lives and odd-even staggering that is not present in the
data. For very neutron-rich isotopes the two models provide very
similar results due to the increase of Q-beta values.  

The right panels compares the present calculation and the the
FRDM+QRPA model with data for nuclei close to the $Z = 50$ closed
proton shell, including the silver and cadmium isotopic
chains. Similar to the two lighter isotopic chains, the D3C*
interaction provides an excellent description of the decay properties
of neutron-rich nuclei, with the results correctly following the trend
along the chain. In the case of silver, the present work reproduces
the trend quite well in comparison with the FRDM. Most likely, because
the FRDM reproduces the half-life of $^{114}$Ag it
cannot provide a good trend for most of the isotopic chain. The half-life of this nucleus is uncharacteristically short compared with the surrounding isotopes due to the $J^{\pi} = 1^{+}$ angular momentum and parity of the ground state which quickly decays into the $0^{+}$ ground state of $^{114}$Cd. For cadmium nuclei both models do quite well, although the FRDM shows unexpected changes in the half-lives along the chain, as in the cases of other elements.

In the following figure, Fig. \ref{fig:ff_chain}, we plot a
comparison of the contribution of the first-forbidden transitions in
the total decay rate, in percent, between the two models. For the
FRDM, we have extracted the values by comparing the results of
Ref.~\cite{Moeller1997} and Ref.~\cite{Moeller2003}. Due to the
changes in the details of the calculation between the two
publications, the values presented here are not precise. We expect
them to be a good qualitative measure of the actual values.  The FRDM
calculation predicts, in two out of four isotopic chains, a large
contribution (more than 50\%) of first-forbidden transitions in nuclei
close to stability, and a decrease in more neutron-rich nuclei. For
nuclei in the isotopic chains of yttrium and silver, FRDM predicts a
steady, small contribution below 20\% for most of the nuclei, with
odd-even staggering. 

The present results predict a rather small contribution of the
forbidden transitions in nuclei close to the stability, with a smooth
increase with additional neutrons. This increase is particularly
visible at the $N = 82$ neutron shell closures for Ag and Cd isotopes,
where the neutrons begin to occupy the $2f_{7/2}$ and the $1h_{9/2}$
orbits, enabling the strong $\nu 2f_{7/2} \to \pi 1g_{7/2}$ and $\nu
1h_{9/2} \to \pi 1g_{9/2}$ transitions, which significantly increase
the contribution of the first-forbidden transitions as these
transitions appear at low excitation energies in the daughter
nucleus. In the two lighter isotopic chains, the contribution is
negligible for the lightest isotopes and increases smoothly up to
$\approx 30$\% where it reaches a plateau. The main difference in the
transitions between e.g. $^{94}$Sr and $^{114}$Sr is the occupied
neutron $2d_{3/2}$ state, which contributes to the total rate by
decaying to the proton $2p$ states. Above the $N = 82$ closed neutron
shell, the contribution of the first-forbidden transitions again rises
with the occupation of the $2f_{7/2}$ and the $1h_{9/2}$ states, just as
in the silver and cadmium isotopes.

As discussed above, our calculations are only strictly applicable to
even-even nuclei. However, we have extended them to odd-A and odd-odd
nuclei in order to provide a complete coverage for r-process
simulations. Our approach cannot account for mismatches between the
angular momenta of the ground states of parent and daughter nucleus in
the decay. However, this aspect becomes less and less important as the
$Q$-value for the decay increases. This is one of the reasons why our
approach becomes better as we move far from stability. Our predictions
for half-lives are certainly not worse that those computed by the
FRDM+QRPA approach~\cite{Moeller2003} that is currently the standard
model for half-lives used in r-process simulations. We will discuss
this aspect in more detail in the next section.

It is also possible to compare the present results with the newest
data available to assess the quality of the extrapolation of
half-lives into currently inaccessible regions.
\begin{figure}[htb]
  \centering
  \includegraphics[width=\linewidth]{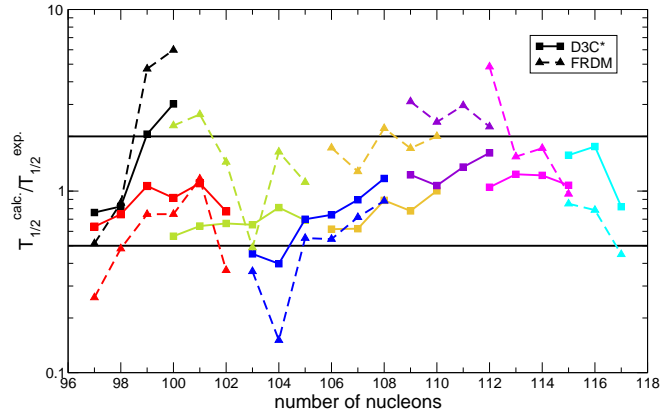}
  \caption{\label{fig:nishimura} (color online) The ratio of the
    calculated and measured half-lives for nuclei in the $36 \le Z \le
    43$ isotopic chains from Ref.~\cite{Nishimura2011}, for D3C*
    interaction (full line) and the FRDM (dashed line). Different
    colors denote the isotopic chains, while the thick black lines
    denote deviations by a factor 2.}
\end{figure}
Fig. \ref{fig:nishimura} compares our calculations with the recently
measured beta-decay half-lives~\cite{Nishimura2011} for a number of
isotopic chains with atomic number in the range $36 \le Z \le 43$. The
results obtained with the D3C* interaction are in agreement with the
experiment within a factor 2. The FRDM+QRPA approach, on the other
hand, tends to predict too long half-lives and displays jumps in the
ratio for specific nuclei. 

\begin{figure}[htb]
\centerline{ 
  \includegraphics[width=\linewidth]{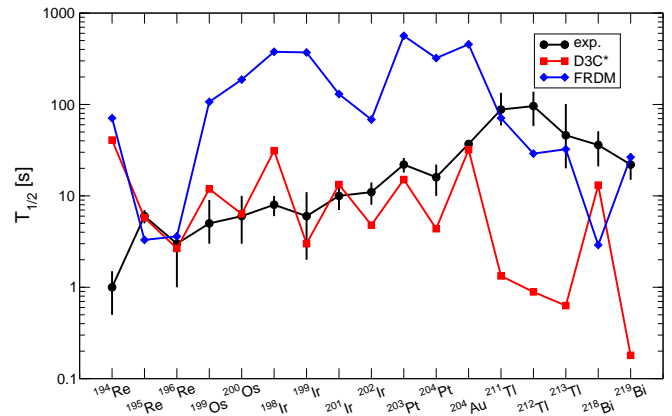}%
}
\caption{\label{fig:hl_heavy} (color online) Comparison of the data
  (denoted with black circles) with the results of the present work
  (red squares) and the FRDM (blue diamonds). }
\end{figure}

As we move to heavier nuclei both models have significant difficulties
while attempting to reproduce the recent data. Fig.~\ref{fig:hl_heavy}
compares the two models with measurements
\cite{Benlliure2012,Benzoni2012}. FRDM significantly overestimates the
half-lives of most of the nuclei, by more than an order of
magnitude. It is only for nuclei around the $Z = 82$ closed proton
shell that provides a reasonable description of the data. An
interesting effect occurs in the case of rhenium isotopes, where the
half-life of $^{194}$Re is overestimated, but half-lives of the two
heavier isotopes are reproduced. The discrepancy seems to be related
to the change of prolate to oblate deformation in moving from
$^{194}$Re to the heavier isotopes~\cite{Moeller1995}. With the
exception of $^{194}$Re the D3C* calculations reproduce all beta-decay
half-lives of the $Z\le 79$ nuclei shown in figure~\ref{fig:hl_heavy}
within an order of magnitude. Larger differences appear for heavier
nuclei. Taking into account that the nuclei shown in
figure~\ref{fig:hl_heavy} have half-lives in the range 1~s to 100~s
one should conclude that both FRDM and D3C* approaches provide a fair
reproduction of data. Certainly, the agreement is worse that the one
obtained in figure~\ref{fig:nishimura} but this should not be consider
as a drawback of the two models discussed to reproduce half-lives of
heavier nuclei. It simply reflects the fact that for heavy nuclei the
half-lives are only known for relatively long lived nuclei close to
the stability that normally have small $Q$-beta decay windows. The
theoretical description of these decays is a challenge for any
theoretical model as the decay rate depends on very few selected
transitions. 

\begin{figure}[htb]
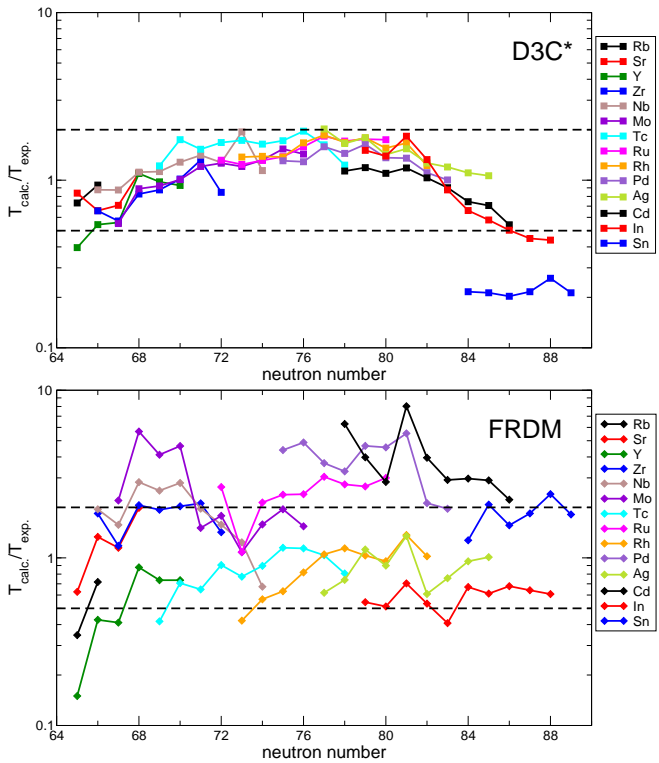

  \centering
  \includegraphics[width=\linewidth]{lorusso_D3CS.eps}\\
  \includegraphics[width=\linewidth]{lorusso_FRDM.eps}%
  \caption{\label{fig:lorusso} (color online) The ratio of the calculated and experimental half-lives for the recently measured 110 isotopes~\cite{Lorusso2015}. In the top panel we present the results obtained in the present study, while in the bottom panel we present the FRDM results~\cite{Moeller2003}.}
\end{figure}

Finally, in Fig.~\ref{fig:lorusso} we compare the theoretical results with the very recent measurement of half-lives of 110 neutron-rich isotopes, 40 of which have not been previously measured. In the upper panel we plot the results of the present study, from the isotopes of rubidium to tin, and from $N = 65$ to $N = 89$. Almost all of the values can be found within a factor of 2 from the data, with very little scatter for different isotopic chains, except for the case of tin where we underestimate the half-lives by a factor of 5. Within a particular chain the data is reproduced smoothly, with very weak odd-even staggering. In the case of FRDM the scatter is much larger, and we also observe some regularities. The half-lives of nuclei in isotopic chains with an even atomic number are systematically overestimated by a factor of two or more - these are the Zr, Mo, Ru, Pd, Cd and Sn isotopes. 

\begin{figure}[htb]
  \centering
  \includegraphics[width=\linewidth]{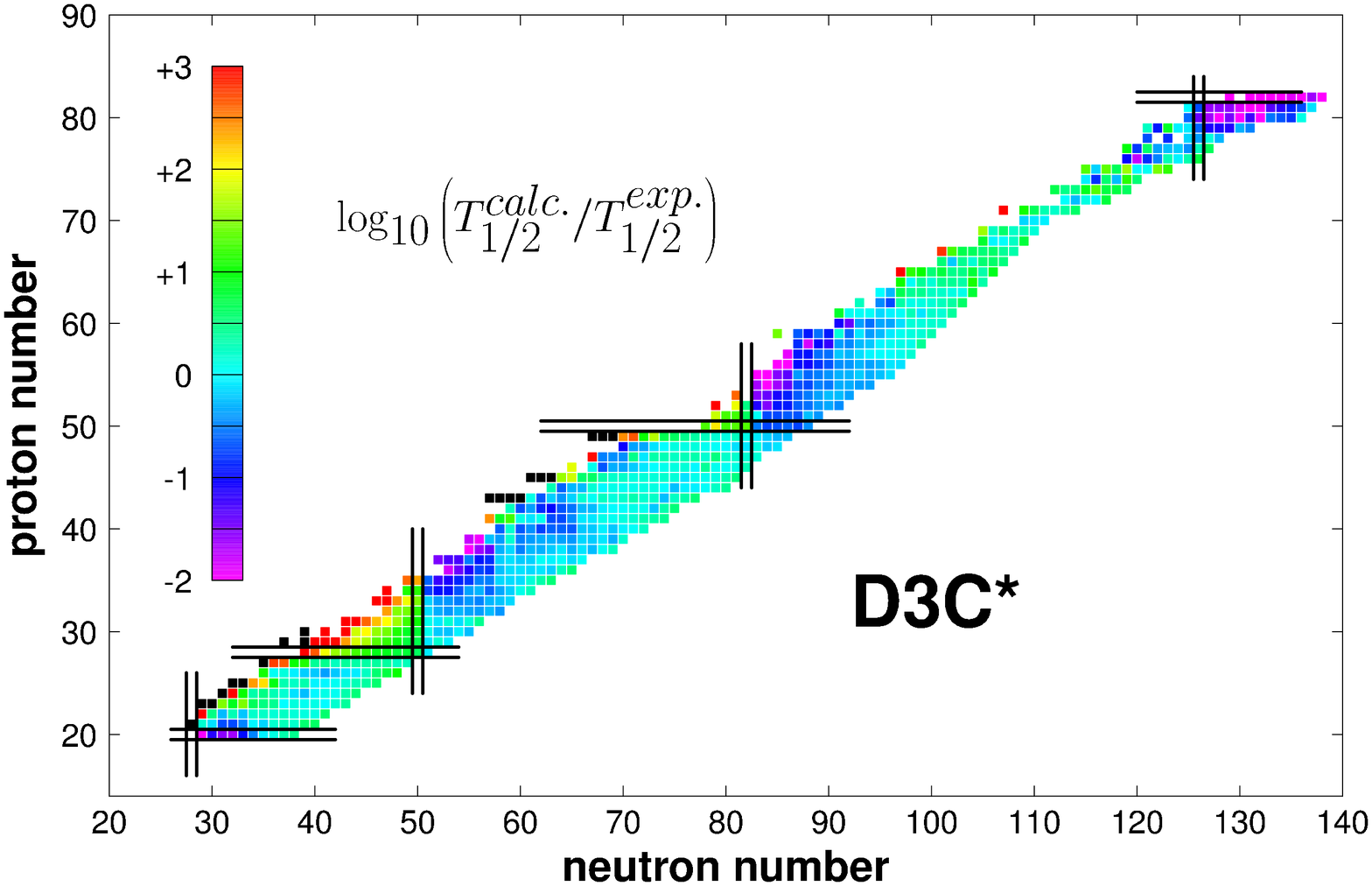}\\
  \includegraphics[width=\linewidth]{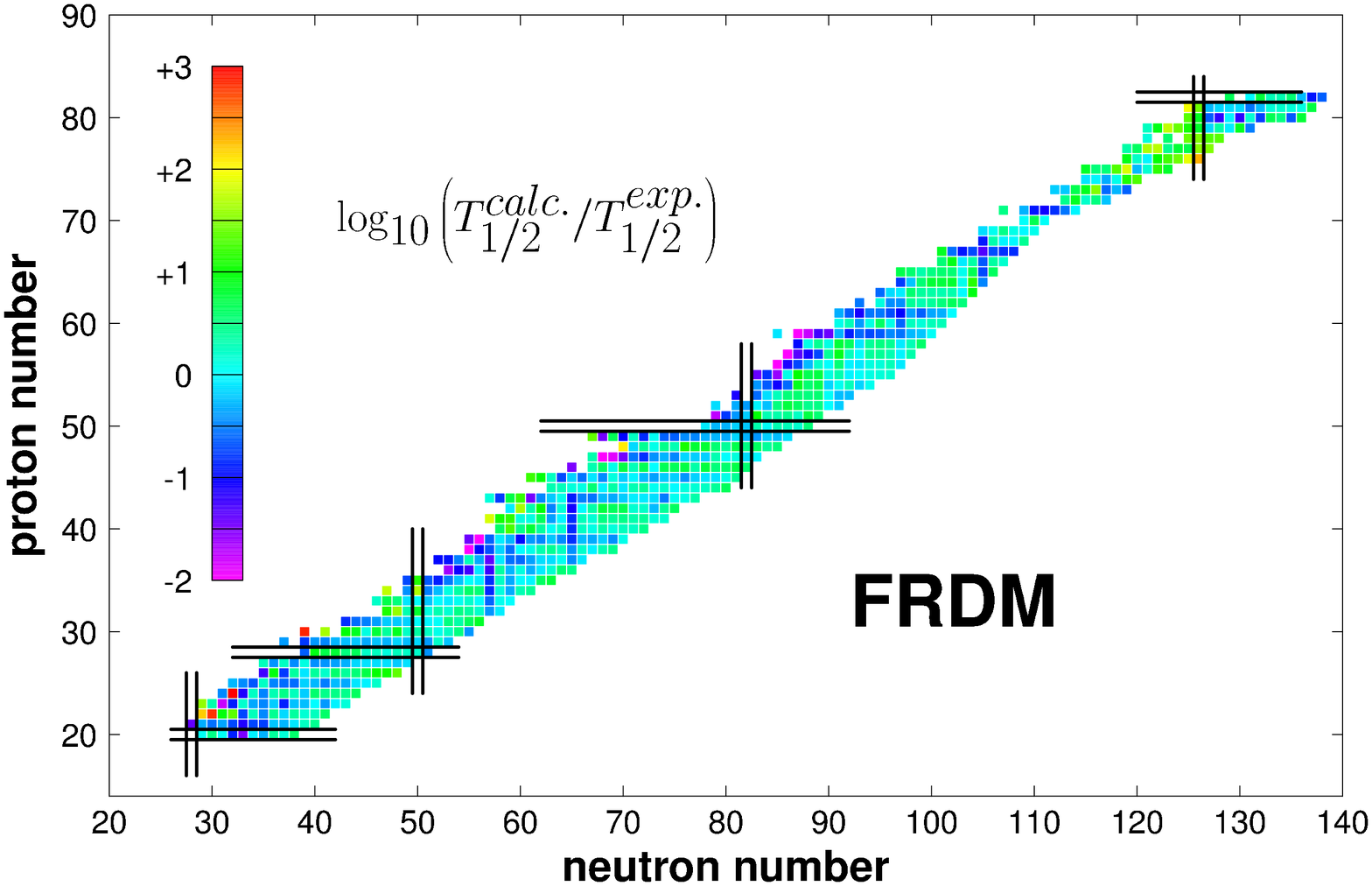}%
  \caption{\label{fig:beta_exp} (color online) The logarithm of the
    ratio of the calculated and experimental half-lives is plotted for
    all neutron-rich nuclei with available data, with half-lives
    shorter than 1 hour. Black squares indicate that the model did not
    provide a finite half-life. In the top panel we plot the results
    of the present study, while in the bottom panel we plot the ratios
    obtained with the FRDM.}
\end{figure}

Fig. \ref{fig:beta_exp} shows the ratio of the calculated and measured
half-lives for neutron-rich nuclei with half-lives shorter than
1~hour. We show the results for the FRDM+QRPA approach and our own
calculations (bottom and top panels, respectively).  The present
calculation has difficulties in reproducing the half-lives for nuclei
close to simultaneous neutron and proton shell closure. In particular,
the greatest discrepancies appear in the regions ``north-west'' of the
doubly closed nuclei $^{78}$Ni and $^{132}$Sn. The amount of low
energy GT strength for such nuclei is very sensitive to
correlations~\cite{Zhi2013} that are not captured by the QRPA
approach. For an improved description of the low-energy strength
distributions of these nuclei, complex configurations need to be
included to shift the transition strength from the GT resonance to low
energies~\cite{Marketin2012,Litvinova2014}. Coupling to collective phonon excitations is known to have a strong impact on the single particle structure around
shell closures~\cite{Litvinova2012}, and will enhance the decay rate of these nuclei~\cite{Niu2015}.

There are also three regions where the D3C* calculations predicts
shorter decay half-lives, i.e. the regions around the $^{90}$Kr
($Z=36$, $N=54$) and $^{140}$Ba ($Z=56$, $N=84$) nuclei, and the $Z
\lesssim 82$ $N>126$ region. The first two cases correspond to regions
of low deformation~\cite{Liu2011}. However, there are indications that
nuclei on that region are gamma
soft~\cite{Bauer.Behrens.ea:2012,Rodriguez:2014}. So far the impact of
gamma-softness on beta-decay rates has not been explored. It is
interesting that the FRDM+QRPA approach also predicts shorter
half-lives for these two regions. In the case of the region $Z
\lesssim 82$ $N>126$, we are dealing with nuclei with half-lives
longer than 100~s. The decay rate is determined by few transitions
that depend very sensitively on the predicted single-particle
structure. In particular, for Hg and Tl isotopes, above the $N = 126$ closed neutron shell the decay is dominated by the Gamow-Teller $\nu 1 i_{11/2} \to \pi 1 i_{13/2}$ and the $1^{-}$ $\nu 1 i_{11/2} \to \pi 1 h_{9/2}$ configurations. These transitions are not possible for isotopes with $N \le 126$, but above the neutron shell closure the $\nu i_{11/2}$ is becoming significantly occupied (occupation probability is already $\approx 0.2$ for $^{210}$Hg and $^{211}$Tl) at an energy of $E_{i_{11/2}} \approx -2.5$ MeV.  

\begin{figure}[htb]
\centering
  \includegraphics[width=\linewidth]{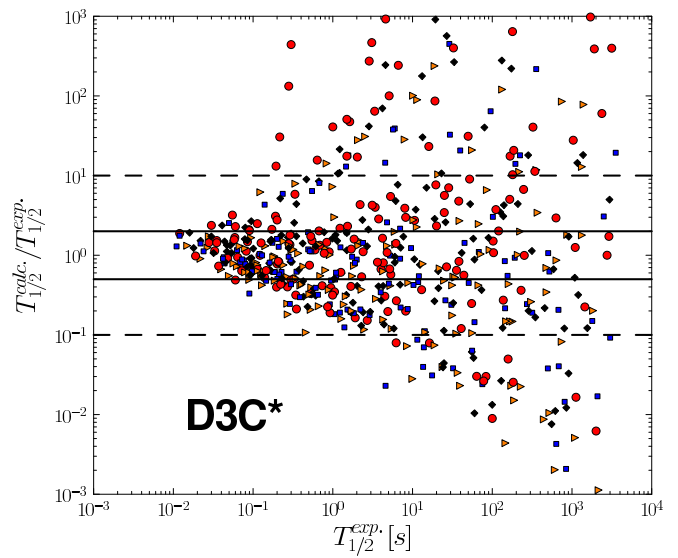}
  \caption{\label{fig:beta_stat} (color online) The ratio of the
    calculated and the experimental half-lives~\cite{Audi2012} versus
    the experimental half-life. Different types of nuclei are denoted
    with different symbols. Even-even nuclei are denoted with blue
    squares, odd-N with orange triangles, odd-Z with black diamonds,
    and odd-odd nuclei are denoted with red circles. }
\end{figure}

Fig. \ref{fig:beta_stat} shows the ratio of the calculated and the
experimental half-lives versus the experimental half-life. Different
symbols are used for even-even, the odd-N, odd-Z and the odd-odd
nuclei. The data is taken from Ref.~\cite{Audi2012}. For long lived
nuclei we observe a very large spread of the ratios due to the high
sensitivity of the decay rate to the details of the low energy
transition strength. Short-lived nuclei have large Q-values and
generally several transitions contribute to the decay rate. This makes
these nuclei less sensitive to the detailed position and strength of
particular transitions. Our results show that the ratio of half-lives
converges towards one with decreasing half-live, for even-even, odd-A
and odd-odd nuclei equally. This behaviour points to the reliability
of the model for all types of nuclei taking part in the r-process, as almost all of the predictions lie within an order of magnitude from the data, with a large majority within a factor of 2. 

\begin{table*}[htb]
  \caption{ \label{tab:stat} Comparison of the average deviations
    (Eq. (\ref{eq:avg})) and their standard deviations
    (Eq. (\ref{eq:stdev})) between the model used in the present work
    and the FRDM. In the upper part of the table, we show the results
    structured with respect to the experimental half-life of the
    nuclei included, while in the lower part of the table we show the
    results for nuclei with half-lives shorter than 1 s, but separated
    into groups of even-even, odd-Z, odd-N and odd-odd nuclei.} 
  \begin{ruledtabular}
    \begin{tabular}{c|dddd}
      & \multicolumn{2}{c}{D3C$^{*}$} & \multicolumn{2}{c}{FRDM} \\
      & \multicolumn{1}{c}{$\bar{r}$} & \multicolumn{1}{c}{$\sigma$} & \multicolumn{1}{c}{$\bar{r}$} & \multicolumn{1}{c}{$\sigma$} \\
      \hline
      $T_{\text{exp.}}$ & & & & \\
      \hline
      $< 1000$ s & 0.011 & 0.889 & 0.021 & 0.660 \\
      $< 100$ s & 0.057 & 0.791 & 0.040 & 0.580 \\
      $< 10$ s & 0.061 & 0.645 & 0.046 & 0.515 \\
      $< 1$ s & 0.011 & 0.436 & 0.019 & 0.409 \\
      $< 0.1$ s & 0.041 & 0.195 & 0.021 & 0.354 \\
      \hline
      nucleus type & & & & \\
      \hline
      even-even & -0.037 & 0.331 & 0.333 & 0.226 \\
      odd-Z & 0.054 & 0.328 & -0.128 & 0.288 \\
      odd-N & -0.086 & 0.387 & 0.124 & 0.436 \\
      odd-odd & 0.089 & 0.582 & -0.179 & 0.409 \\
      total & 0.011 & 0.436 & 0.019 & 0.409 \\
    \end{tabular}
  \end{ruledtabular}
\end{table*}

In Table \ref{tab:stat} we provide a quantitative analysis of the
results, by using several measures of the model precision. Because
both the observable (half-life) and the ratio of the calculated and
experimental values span orders of magnitude we define $r_{i}$ as the
logarithm of the ratio of calculated and experimental half-lives
\begin{equation}
r_{i} = \log_{10} \frac{T_{1/2}^{\text{calc.}}}{T_{1/2}^{\text{exp.}}},
\end{equation}
with the average value of the ratio providing a measure of the global
deviation from experimental values 
\begin{equation} \label{eq:avg}
\bar{r} = \frac{1}{N} \sum_{i=1}^{N} r_{i}, 
\end{equation}
while the standard deviation provides a measure of the
spread~\cite{Moeller1990} 
\begin{equation} \label{eq:stdev}
\sigma = \left[ \frac{1}{N} \sum_{i=1}^{N} \left(r_{i} - \bar{r}
  \right)^{2} \right]^{1/2}. 
\end{equation}
The average deviation and the spread obtained using the D3C$^{*}$ and
the FRDM model~\cite{Moeller2003} are shown. The top part of the table
provides data for nuclei with the experimental half-life shorter than
a particular value. The average value of deviation from experiment
$\bar{r}$ is comparable between the two models. There are small
differences for particular sets of nuclei, but they are not
significant. The resulting standard deviation is larger with the
D3C$^{*}$ for long-lived nuclei, confirming that the description of
nuclei with small $Q_{\beta}$ values is challenging for the model. For
nuclei with half-lives shorter than 1 s, the models are comparable
both in the average and the standard deviation, and for even less
stable nuclei D3C$^{*}$ even manages to provide a smaller spread
of ratios.

In the bottom part of the table we provide the results, including
nuclei with the experimental half-life lower or equal to 1 s, but
shown for each type of nuclei, i.e. even-even, odd-Z, odd-N and
odd-odd, separately. From this point of view, the results of the two
models are significantly different. In the present work, the
half-lives of even-even and odd-N nuclei are somewhat underestimated
by roughly 10\% - 20\%, and the half-lives of odd-Z and odd-odd nuclei
are, on average, overestimated by 15\% and 20\%, respectively. These
results agree with the assumption that the model is capable of
describing all types of nuclei equally well, as the half-lives of
odd-A and odd-odd nuclei are reproduced as well as half-lives of
even-even nuclei. The difference between various types of nuclei
becomes more evident by looking at the standard deviations. Here, the
spread of values for even-even and odd-A nuclei is comparable, but the
standard deviation of the results for odd-odd nuclei is almost twice
as large, pointing to potential angular momentum issues in the ground state of an odd-odd nucleus. It is interesting that our model does particularly well for
even-N nuclei that are the most relevant nuclei for r-process
nucleosynthesis as these nuclei are favored by neutron-captures. 

The FRDM+QRPA half-lives show a distinctly different behaviour. As
opposed to the D3C$^{*}$ model, the FRDM overestimates the half-lives
of even-even and odd-N nuclei, by a factor of 2.15 and 1.33,
respectively. That the predicted half-lives of even-even nuclei are on
average twice the experimental value is surprising as, similarly to
our calculation, the model should be more suitable
for these nuclei. This fact may have important consequences for
r-process nucleosynthesis and is might be responsible for the
differences in r-process nucleosynthesis discussed on
section~\ref{sec:cons-heavy-elem}. However, waiting point nuclei also include odd-Z, even-N nuclei, and in both models there is a cancellation between even-even and odd-Z nuclei. 

On the other hand, the half-lives of odd-Z and odd-odd nuclei are underestimated by
approximately 30\% and 50\%, respectively. The standard deviations are
comparable between the two models for odd-A nuclei, but are smaller
for even-even and especially for odd-odd nuclei. Still, for both
models the odd-N and odd-odd nuclei are the most difficult to describe
and result with the largest spreads. So, even though the average
$\bar{r}$ of the FRDM model is close to zero, it is actually a result
of the cancellation of overestimated and underestimated half-lives in
different types of nuclei. This behaviour has also appeared in
previous evaluations of the $\beta$-decay rates (see
Ref.~\cite{Moeller1997}, Table B.), where the half-lives of
short-lived even-even nuclei were, on average, overestimated by almost
a factor of 4, while the half-lives of odd-A and odd-odd nuclei were
underestimated by a factor of 2. In comparison, the results of the
present work are more consistently close to reproducing the data, even
though the larger standard deviations highlight some unresolved
issues.

\begin{figure}[htb]
  \centering
  \includegraphics[width=\linewidth]{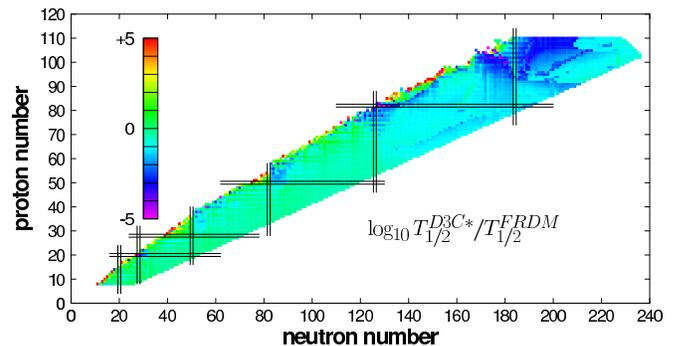}%
  \caption{\label{fig:beta_comparison} (color online) The ratio of
    halflives obtained by the two models.}  
\end{figure}

Finally, in Fig.~\ref{fig:beta_comparison} we plot the ratio of
half-lives obtained in the present study with the FRDM
half-lives~\cite{Moeller2003}. For light and medium-heavy nuclei, the
results are quite similar. Close to the valley of stability, the
D3C$^{*}$ model tends to provide longer half-lives than the FRDM for
particular nuclei, especially so in the regions ``north-west'' of the
doubly closed nuclei $^{78}$Ni and $^{132}$Sn. Further from the valley
of stability the two models provide comparable results, with the
difference that the present calculation predicts smoother increase in
the decay rates. For very heavy nuclei with $N > 126$ however, the
present study predicts half-lives to be shorted by more than an order
of magnitude than the FRDM predictions. This is especially clear in
nuclei with high atomic numbers, i.e. $Z \gtrapprox 95$ where the
difference is as large as three orders of magnitude. This may have
significant consequences on the dynamics of the r-process
nucleosynthesis in neutron star merger
conditions~\cite{Eichler.Arcones.ea:2014}

\subsection{Impact of First-forbidden transitions}

The results presented in this manuscript are the first global calculations of beta-decay half-lives for neutron-rich nuclei that treat Gamow-Teller and first-forbidden transitions on an equal footing. In this section, we provide a detailed analysis on the impact of first-forbidden transitions with particular focus on nuclei with magic neutron numbers $N=50, 82$ and 126 that have been the subject of many theoretical calculations. Additionally, for the $N=50$ and 82 experimental data is available that has contributed to constrain the calculations. 

\begin{figure}[htb]
  \centering
  \includegraphics[width=\linewidth]{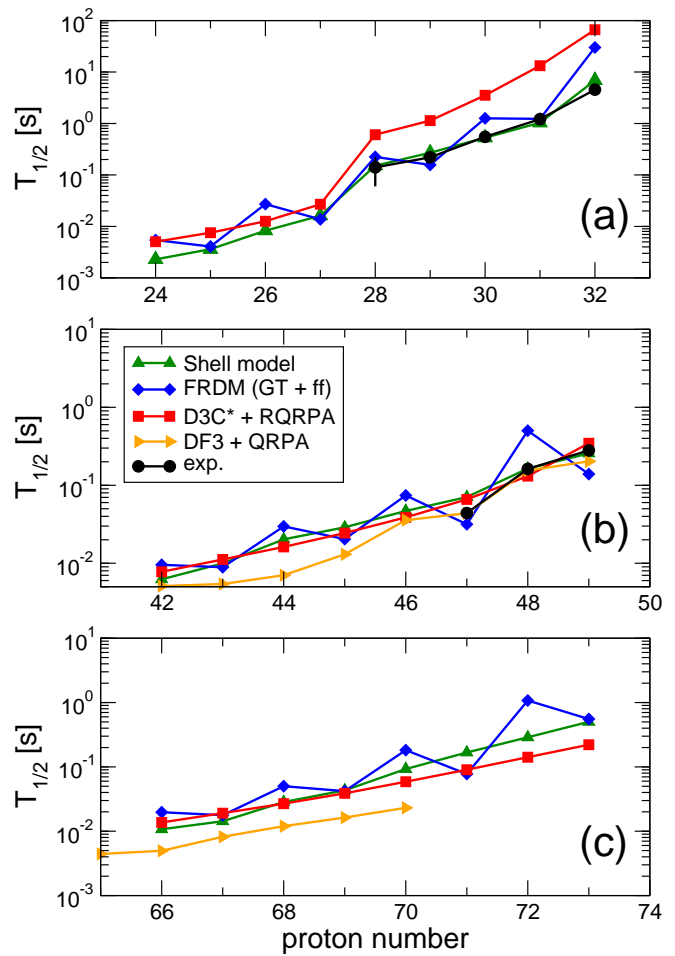}
  \caption{(color online) Beta-decay half-lives for $N=50$ (a), $N=82$
    (b), and $N=126$ (c) isotopic chains. The present results are compared
    with those of FRDM~\cite{Moeller2003}, the
    shell-model~\cite{Zhi2013}, DF3~\cite{Borzov2003a} and
    data~\cite{Audi2012}. \label{fig:nmagiclives} }
\end{figure}

\begin{figure}[htb]
  \centering
  \includegraphics[width=\linewidth]{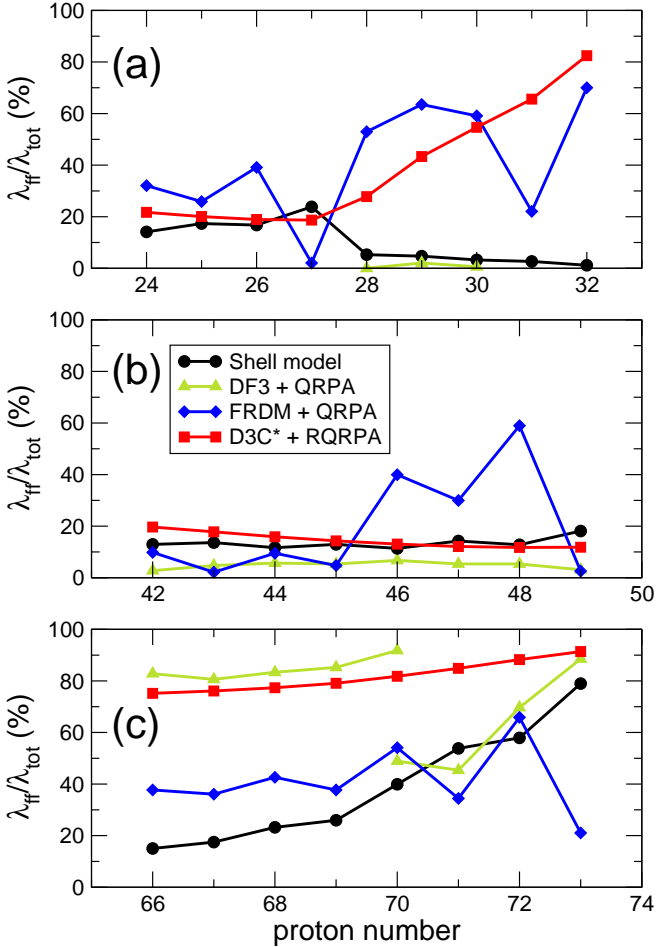}%
  \caption{\label{fig:ff_perc} (color online) The contribution of the
    first-forbidden transitions in the total decay rate for the $N =
    50$, $N = 82$ and $N = 126$ isotonic chains, in the top, central
    and bottom panels, respectively. Present results, denoted with
    black squares, are compared with the FRDM
    calculation~\cite{Moeller2003} (denoted with green triangles), DF3
    calculation~\cite{Borzov2006} (denoted with blue diamonds) and a
    recent shell model calculation~\cite{Zhi2013} (denoted with red
    circles).} 
\end{figure}

Fig.~\ref{fig:nmagiclives} compares the beta-decay half-lives for
r-process nuclei with neutron magic number $N=50, 82$ and 126 with the
FRDM+QRPA model~\cite{Moeller2003}, the shell-model calculations of
ref.~\cite{Zhi2013} and data~\cite{Audi2012}. For the $N=50$ isotones
(upper panel) with $Z\geq 28$ the present calculations fail to
reproduce the measured half-lives. In these nuclei the $f_{7/2}$
proton shell is fully occupied suppressing low energy Gamow-Teller
transitions. The decay consequently proceeds mainly by first-forbidden
transitions resulting in long decay half-lives. As a consequence the
model predicts a large contribution of forbidden transitions to the
decay rate (see fig.~\ref{fig:ff_perc}). This is probably a limitation
of the QRPA type calculations that are not able to produce enough
correlations around the proton magic number $Z=28$. This problem does
not seem to be present in the FRDM+QRPA approach. However, one should
kept in mind that in this approach the Gamow-Teller and
first-forbidden contributions are not derived based on the same
microscopic model. As discussed above the FRDM+QRPA approach tends to
produce strong odd-even effects that are not present in the data nor
in the other approaches.

The situation is different for the $N=82$ and $N=126$ isotonic
chains. In these nuclei there is no proton shell closure that hinders
low energy Gamow-Teller transitions and in principle both Gamow-Teller
and forbidden transitions are possible. The present study predicts a
very similar contribution of first-forbidden transitions to the
shell-model, $\sim 20$\%, for all the $N=82$ isotones. FRDM+QRPA
predicts rather large forbidden contributions for $Z=46,47,48$ and
similar values for the other isotones. The odd-even staggering present
in the beta-decay half-lives is also manifested on the contribution of
forbidden transitions. For $N=126$ all models predict an enhanced
contribution of forbidden transitions. Ref.~\cite{Zhi2013} argues
that the contribution of forbidden transition should decrease with
decreasing proton number. This reduces the number of protons in the
$h_{11/2}$ increasing the role of $\nu h_{9/2} \rightarrow \pi
h_{11/2}$ Gamow-Teller transitions. While all models predict such a
reduction, the magnitude varies from model to model. The present
calculations predict a relatively minor reduction from 90\% at $Z=73$
to 70\% at $Z=66$, while the shell-model predicts a much larger
reduction reaching below 20\% at $Z=66$. Experimental data will be
very important to further constrain the theoretical calculations. 

\begin{figure}[htb]
  \centering
  \includegraphics[width=\linewidth]{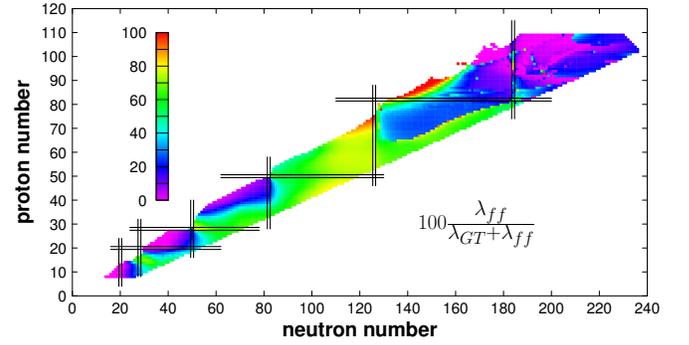}%
  \caption{\label{fig:ff_total} (color online) The contribution of the
    first-forbidden transitions in the total decay rate for nuclei
    with predicted half-lives shorter than 1~s. Black lines indicate
    the position of (predicted) closed proton and neutron shells.}
\end{figure}

Fig~\ref{fig:ff_total} shows the contribution of beta-decay half-lives
for all nuclei with predicted beta-decay half-lives shorter than
1~s. In general, first-forbidden transitions do not noticeably impact
the total decay rate in nuclei close to the valley of stability with
$Z \le 50$. Here, protons and neutrons fill orbits with the same
parity and the decay is dominated by Gamow-Teller transitions. It is
interesting to notice the similar behaviour for nuclei located around
$^{78}$Ni and $^{132}$Sn. Moving in the ``south-west'' direction
($Z<28$, $N<50$ and $Z<50$, $N<82$ respectively) forbidden transitions
are initially suppressed. However, once protons start to occupy orbits
with different parity ($Z<20$ or $Z\lesssim 40$) the contribution of
forbidden transitions grows to values of 40\% to 60\% depending on the
detailed nuclear structure. Moving in the ``east'' direction ($N>50$
and $N>82$ respectively), we see that the contribution of forbidden
transitions is rather independent of the charge number and does not
vary when crossing the proton shell closure at $Z=28$ and $Z=50$,
respectively. In these regions, for neutrons and protons there are
valence orbits with both positive and negative parity and consequently
Gamow-Teller and forbidden transitions contribute in similar amounts
to the decay. The situation is different in the region between $N=50$
and $N=82$. Here, once protons completely fill the $pf$ shell ($Z>40$)
forbidden transitions are suppressed and the decay is dominated by
Gamow-Teller transitions. 

An interesting phenomena occurs once we move to nuclei with $N>126$
with a clearer manifestation once we reach $N=184$. For these nuclei,
neutrons and protons occupy orbits that differ in two units of the
main oscillator quantum number, $\mathcal{N}= 2 n +l$. Under the
assumption of isospin symmetry the single particle orbits for
neutrons and protons will be identical and consequently Gamow-Teller
transitions will be exactly zero. However, due to isospin breaking
mainly due to the coulomb interaction the proton and neutron single
particle states are not identical allowing for Gamow-Teller
transitions, even if somewhat suppressed. This is the reason why
forbidden transitions dominate in this whole region with GT
transitions representing around 20\% of the decay rate. 

\subsection{$\beta$-delayed neutron emission}

$\beta$-delayed neutron emission is another important component in the
late stages of r-process nucleosynthesis. Here we approximate the
probability of emission of x neutrons as the ratio of the rates
between $S_{xn}$ and $S_{(x+1)n}$ separation energies to the total
decay rate, i.e.
\begin{equation} \label{eq:emission_prob}
P_{xn} = \frac{\displaystyle \sum_{i,E_{i} = S_{xn}}^{\min (Q_{\beta},S_{(x+1)n})} \lambda_{i}}{\sum_{i} \lambda_{i}}.
\end{equation}
Neutron emission probabilities are, therefore, very sensitive to the
neutron separation energies and display strong odd-even effects. As
mean-field models cannot reproduce this staggering we employed the
global nuclear mass model~\cite{Liu2011} to obtain the neutron
separation energies. Additionally, only for the purpose of calculating
the emission probabilities, we smeared the beta strength distributions
with a Lorentzian in order to reduce the sensitivity of
the neutron emission probabilities on the positions of particular
transitions. The width of the Lorentzian was determined by calculating the average number of delayed neutrons emitted by the fission fragments of $^{325}$U, $\left\langle n \right\rangle = 0.0158 \pm 0.0005$~\cite{Keepin1957}. In this way the width was adjusted to $\Gamma = 65$ keV. From the probabilities, we obtain the average number of emitted neutrons after the decay of a nucleus as 
\begin{equation}
\left\langle n \right\rangle = \sum_{i} iP_{in}.
\end{equation}

\begin{figure}[htb]
\centering
  \includegraphics[width=\linewidth]{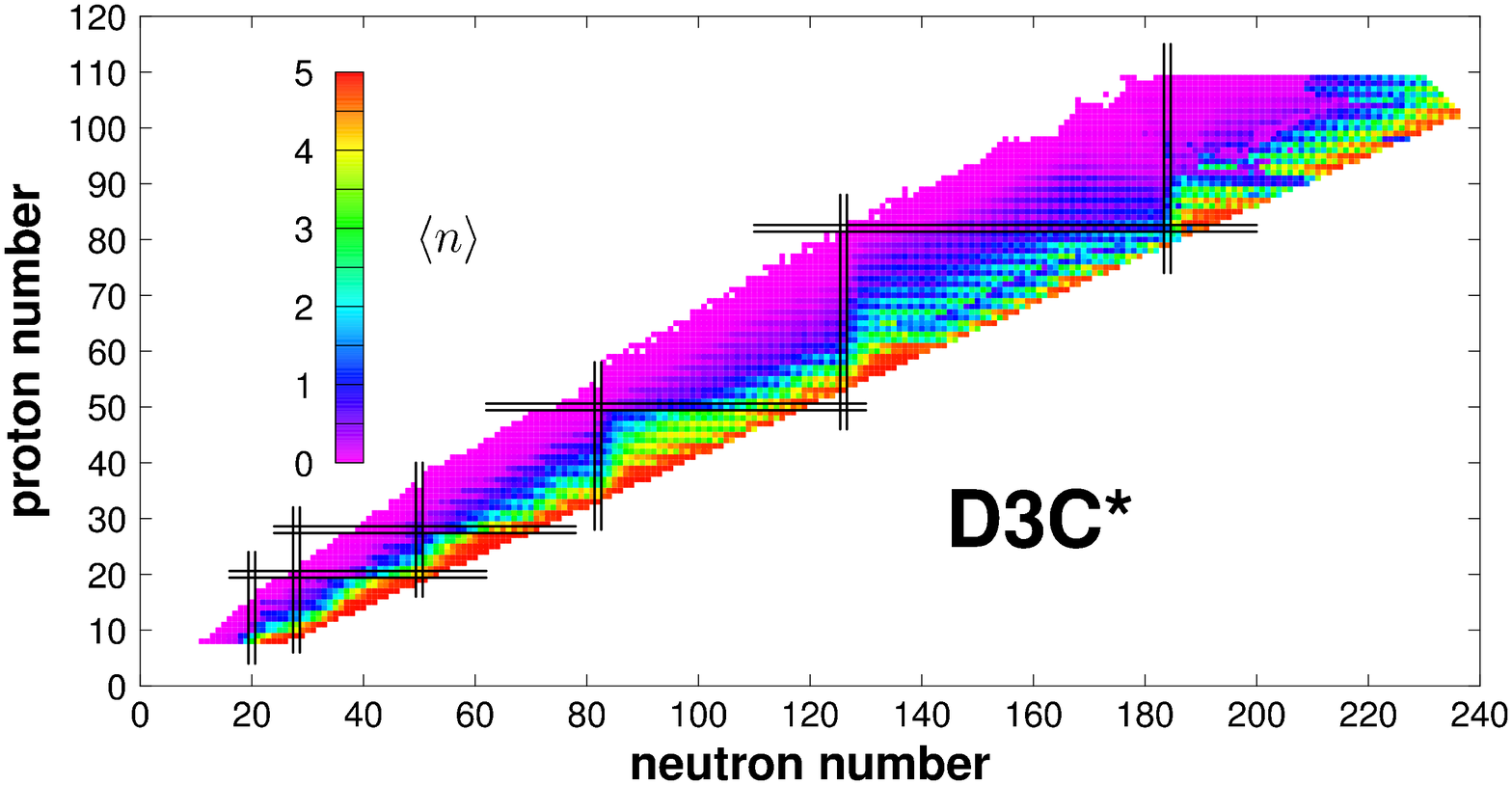}\\
  \includegraphics[width=\linewidth]{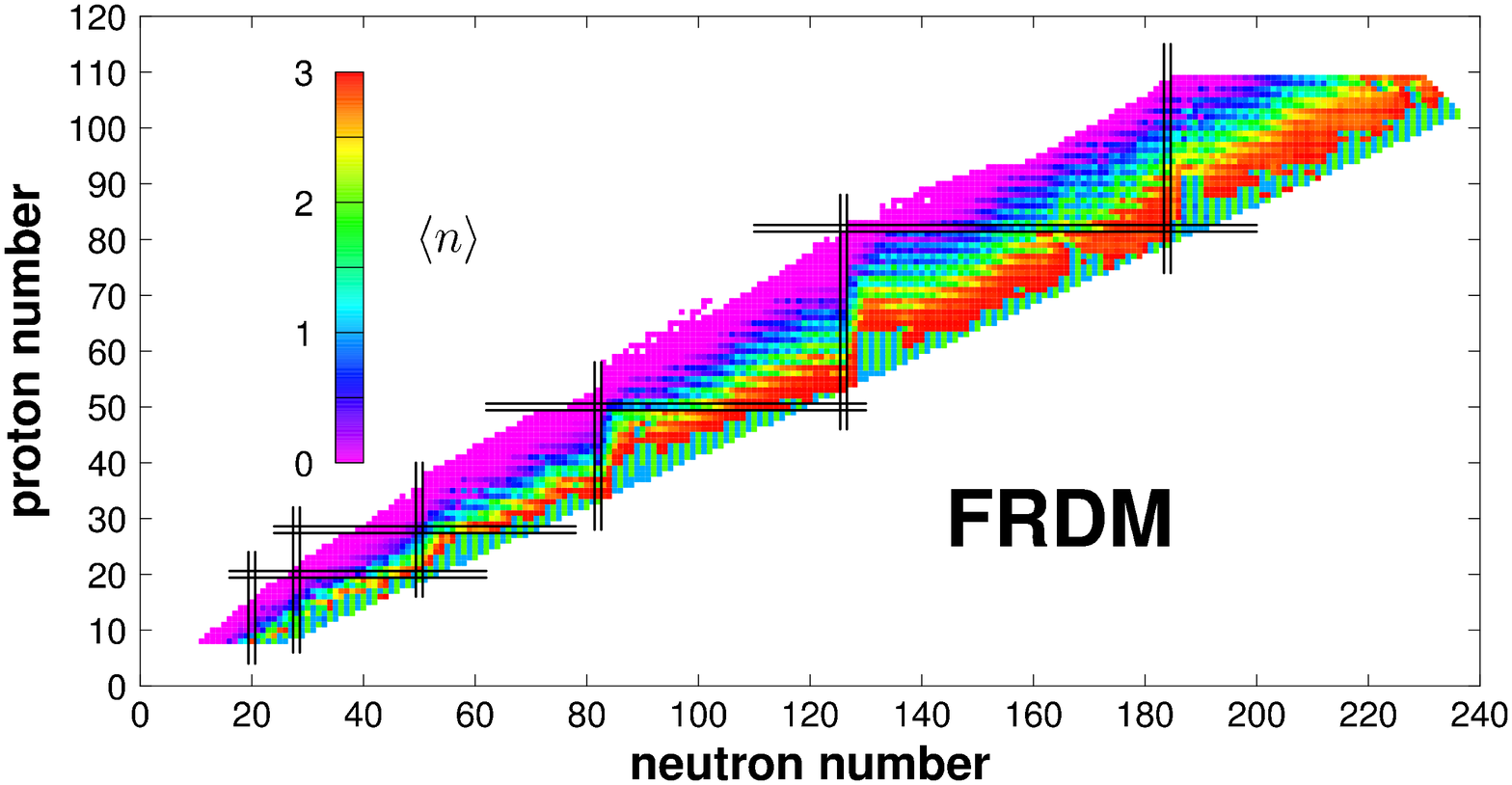}%
  \caption{\label{fig:emission} (color online) Average number of
    emitted neutrons after $\beta$-decay calculated with the
    FRDM and the D3C$^{*}$ models. Note that because the available FRDM results only contain the emission probabilities for up to 3 neutrons emitted, and in the present work the probabilities are obtained for up to 5 emitted neutrons, the colour scales are different.} 
\end{figure}

In Fig. \ref{fig:emission} we plot the average number of emitted
neutrons for neutron-rich nuclei included in this study. Close to the
valley of stability the Q-values are smaller than the one neutron
separation energies, making it impossible for the daughter nucleus to
de-excite via neutron emission. With additional neutrons both the
Q-value increases and the separation energies decrease enabling the
emission of one, two or more neutrons after decay. This is
particularly evident above neutron shell closures where the average
number of emitted neutrons increases due to a drop in the separation
energies. Another characteristic feature of $\left\langle n
\right\rangle$ is the odd-even staggering as a function of atomic
number. Isotopic chains with an even number of protons typically have
a smaller average number of emitted neutrons, i.e. smaller
probabilities of emitting more neutrons, than their odd-proton
neighbouring chains which mimics the behaviour of the Q-values. A
smaller odd-even staggering is also present within any isotopic chain
as a consequence of the staggering of the neutron separation energies.

While both the D3C$^{*}$ and the FRDM models reproduce
these general features, in other aspects there are significant
differences. Firstly, the FRDM evaluation was limited to a maximum of
3 emitted neutrons per decay, while the present study takes into
account the possibility for up to 5 neutrons to be emitted in a
decay. The consequence of this change is evident in very neutron-rich
nuclei where the neutron separation energies are very small, and where
the D3C* model predicts a larger average number of emitted neutrons
which continuously increases up to the drip-line, while the FRDM model
predicts a saturation of $\left\langle n \right\rangle$. On the other
hand, for isotopes closer to stability the D3C$^{*}$ model predicts
smaller vales of $\left\langle n \right\rangle$ and a more gradual
increase towards the neutron drip-line. Finally, in very neutron-rich
nuclei the FRDM displays decidedly unphysical, sudden oscillations
between the values of $\left\langle n \right\rangle = 1$ and
$\left\langle n \right\rangle = 2$ while the present study predicts a
smooth increase with the neutron number. 

\subsection{Consequences for heavy-element nucleosynthesis}
\label{sec:cons-heavy-elem}

To explore the impact of the obtained results on heavy element
nucleosynthesis, we have performed r-process nucleosynthesis
calculations based on the ``hot'' and ``cold'' r-process conditions
from ref.~\cite{Arcones2011}. In the ``hot'' r-process scenario the
r-process evolves under an $(n,\gamma)\rightleftarrows(\gamma,n)$
equilibrium that breaks down once the r-process freeze-out is
reached. For the ``cold'' r-process scenario the evolutions proceeds
by a competition between neutron-captures and beta-decays. The neutron
capture and photodissociation rates used on the network calculations
are based on the statistical model
approach~\cite{Rauscher.Thielemann:2000} using the FRDM mass
model~\cite{Moeller1995}. The only difference between the calculations
is the use of the FRDM+QRPA beta-decay half-lives and beta-delayed neutron emission probabilities~\cite{Moeller2003} or those computed in the present study. 

\begin{figure}[htb]
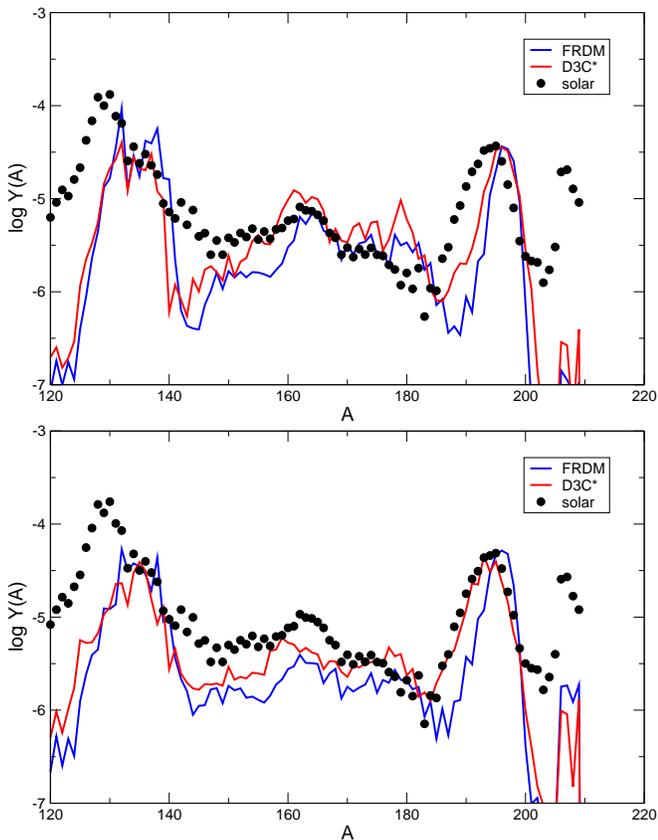

  \centering
  \includegraphics[width=\linewidth]{Fig13a.eps}\\
  \includegraphics[width=\linewidth]{Fig13b.eps}
  \caption{\label{fig:rabund} (color online) The abundances of heavy
    nuclei resulting from a hot r-process (upper panel) or a cold
    r-process (lower panel) calculation using the half-lives obtained
    from the FRDM and the current model. The solid circles denote the
    solar r-process abundances.}
\end{figure}

\begin{figure}[htb]
  \centering
  \includegraphics[width=\linewidth]{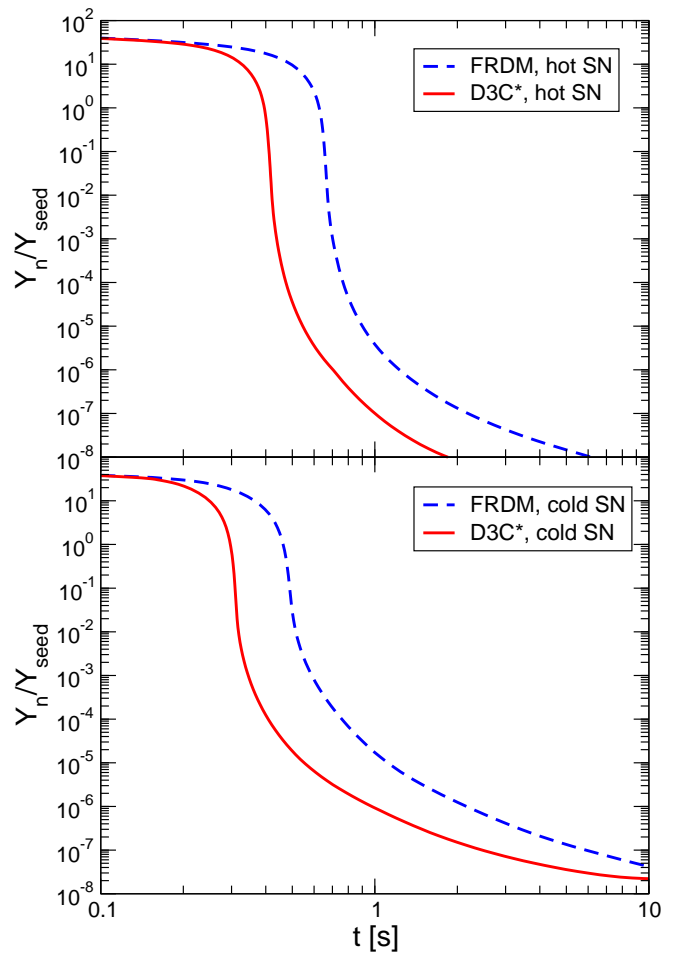}%
  \caption{\label{fig:nseed} (color online) The evolution of the
    neutron-to-seed ratio during the r-process. The upper panel shows
    the results for the hot r-process while the lower panel for the
    cold r-process. Dashed lines denote the results obtained with the
    FRDM decay data, and full lines denote results obtained in the
    present study.} 
\end{figure}

In Fig.~\ref{fig:rabund} we compare the abundances obtained with the
new decay half-lives and delayed neutron emission probabilities with
the results obtained using the FRDM values for both the hot and cold
r-process trajectories. One can notice important differences
particularly in the region of the third r-process peak, $A\sim 195$
between the calculations. There are also differences in the abundances
around the second peak but the astrophysical conditions explored by
the present trajectories are expected to contribute mainly to the
third r-process peak region, see ref.~\cite{Arcones2011}.  To get
further insight in the origin of the differences in the abundances we
show in figure~\ref{fig:nseed} the evolution of the neutron-to-seed
ratio in the present calculations. We clearly see that for those
calculations that use the D3C* the r-process freeze-out, defined as
the moment when the neutron-to-seed ratio reaches a value of 1,
occurs at an earlier time. An earlier freeze-out has important
consequences for the shaping of the final abundance pattern during the
decay to stability. This is because it occurs at a higher value of
the density that speeds up the freeze-out producing a faster and
larger drop in the neutron-to-seed ratio. As a consequence, there is
less time for neutron-captures during the freeze-out. These neutron
captures are in fact responsible for shifting the peaks to higher $A$
values as discussed in ref.~\cite{Arcones2011}. 

\begin{figure}[htb]
  \centering
  \includegraphics[width=\linewidth]{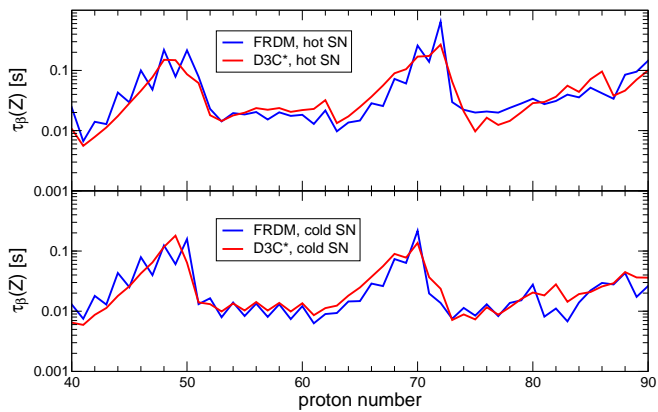}%
  \caption{\label{fig:lambdaz} (color online) The isotopic lifetimes
    defined in Eq. (\ref{eq:lambdaz}) as a function of the atomic
    number, at the moment of r-process freze-out. In the top panel we
    show the results for the hot r-process, and in the bottom panel
    the results for the cold r-process.}
\end{figure}

The evolution proceeds faster with the new half-lives because they
tend to be shorter than those computed by the FRDM+QRPA approach. This
is particularly the case near to the magic shell closures $N=82$ and
$N=126$ where the r-process moves closer to the stability and reaches
the longest living nuclei. Both for the hot and cold r-process
conditions, the r-process has reached beta-flow equilibrium by the
freeze-out time. Under these conditions the product of the average
beta-decay rate times the abundance for each isotopic chain is
constant, i.e. $\lambda_\beta(Z) Y(Z)= \mathrm{constant}$,  with
$Y(Z)=\sum_A Y(Z,A)$ and  

\begin{equation} \label{eq:lambdaz}
\lambda_{\beta}(Z)= \frac{1}{\tau_\beta(Z)} = \frac{1}{Y(Z)} \sum_{A}
\lambda_{\beta}(Z,A) Y(Z,A).
\end{equation}
This means that the isotopic abundances, $Y(Z)$, are proportional to
the lifetime, $\tau(Z)$, and the material accumulates in regions with
the longest lifetimes that correspond to the magic shell
closures. Figure~\ref{fig:lambdaz}, shows the isotopic lifetimes at
the freeze-out for the different calculations. One can clearly see
that the lifetimes are substantially longer for the FRDM+QRPA approach
particularly close to the magic neutron numbers $N=82$, corresponding
to $Z\sim 50$, and $N=126$, corresponding to $Z\sim 70$.

We expect that the variations discussed above will remain for a broad
range of r-process conditions and in particular also for neutron-star
merger conditions. This aspect will be explored in more detail in a
future publication. An important aspect for r-process calculations
under r-process conditions is the energy generation during the
r-process. This depends mainly on the beta-decay rates and determines
the evolution of temperature during the
r-process~\cite{Mendoza-Temis.Martinez-Pinedo.ea:2014}. The nuclear energy
production is also important to determine the evolution of material
ejected in highly eccentric loosely bound orbits. Depending on the
energy produced by the r-process this material can become unbound
affecting the late time evolution of short gamma-ray burst afterglows
assuming they are powered by matter
fallback~\cite{Metzger.Arcones.ea:2010}. At later times the
radioactive heating from the beta-decay of r-process material will
also affect the neutron-star merger remnant
evolution~\cite{Rosswog.Korobkin.ea:2014} and the electromagnetic
transient, kilonova,
lightcurve~\cite{Metzger.Martinez-Pinedo.ea:2010,Roberts.Kasen.ea:2011,Bauswein.Goriely.Janka:2013}. For all these applications it is important to know how much of the energy
produced by beta-decay is lost in the form of neutrinos and how the remaining
energy is distributed between the electrons and gamma-rays emitted from
nuclear deexcitation. We have determined for each decay these
quantities based on the computed beta-decay strengths. A small part of the energy will be
carried away by the neutrons emitted after the beta-decay. Our gamma
channel does in fact also include this energy as it is assumed that
the emitted neutrons will thermalize in a relatively short
timescale. All this information is included in the supplemental material
that also includes the relevant beta-decay rates. 

\begin{figure}[htb]
  \centering
  \includegraphics[width=\linewidth]{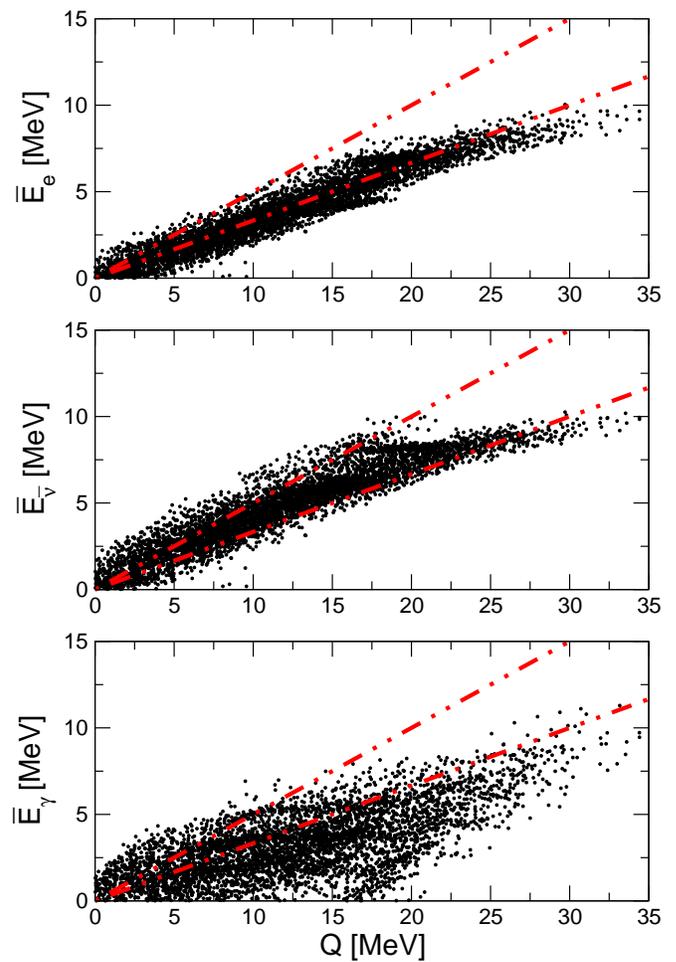}
  \caption{(color online) Average energy of the electrons, neutrinos and photons
    produced after beta-decay \label{fig:energy}}
\end{figure}

Figure~\ref{fig:energy} shows the average energy of electrons,
neutrinos and photons produced after beta-decay as a function of the
decay $Q$-value. The red lines denote the one-half and one-third ratios between the energy and the Q-values. 
\begin{equation}
\bar{E}_{e,\bar{\nu},\gamma}^{1}(Q) = \frac{1}{2} Q, \quad \bar{E}_{e,\bar{\nu},\gamma}^{2}(Q) = \frac{1}{3} Q.
\end{equation}
One can see that the ratio of the average electron energy and the nuclear $Q$-value is concentrated around one third, with the same artio for the antineutrinos is in the range 0.3-0.5. r-process nucleosynthesis is not expected to be sensitive to this range of variations in the heating rate~\cite{Mendoza-Temis.Martinez-Pinedo.ea:2014}. However, kilonova light curves are more sensitive to this range of variation~\cite{Metzger.Martinez-Pinedo.ea:2010} and our calculations are the first to provide this information for future modeling.

\section{\label{sec:conclusion} Conclusion and outlook}

In this study we have performed a large scale calculation of
$\beta$-decay rates and $\beta$-delayed neutron emission probabilities
for neutron-rich nuclei with $8 \le Z \le 110$. A fully
self-consistent theoretical framework was employed based on the
relativistic nuclear energy density functional. The advantages of this
approach are: (i) self-consistent modeling of all relevant transition
matrix elements without any adjustment of the model parameters to the
nucleus under consideration, and (ii) equal treatment of allowed and
first-forbidden transitions. The r-process involves even-even nuclei,
but also odd-A and odd-nuclei. It is, therefore, very important that
all types of nuclei are treated within a single theoretical
framework. We have tested the quality of the description of odd nuclei
by comparing the $\beta$-decay $Q$-values with those of the FRDM and
the data for the silver and cadmium isotopic chains, where the results
show very good agreement with the available data. Additionally, in
agreement with data there is no noticeable odd-even staggering in the
obtained half-lives, and the statistical analysis shows only a small
tendency to overestimate the half-lives of odd-Z and odd-odd
nuclei. Additionally, even though the model used assumes a spherical
shape of the nucleus under consideration, it has been shown that it
can accurately describe the reaction rates both in the regions of the
nuclear chart close to closed shells and in open-shell nuclei known to
be well-deformed. Thus, the model employed is shown to reproduce
$\beta$-decay data across the whole neutron-rich side of the nuclear
chart with good precision, regardless of the type of nucleus.

In general, the key features of the decay rates across the nuclear
chart have been reproduced. We obtain a very good agreement with the
data for isotopic chains with both an even and odd atomic number,
without any noticeable odd-even staggering along the chain and closely
following the trend of the measurements. This characteristic ensures a
smoother behaviour during the final decay towards the line of
stability in the late stages of the heavy element nucleosynthesis. The
FRDM results, however, often displays erratic jumps in the half-lives
without an underlying physical reason, as is evident in e.g. strontium
and cadmium isotopic chains. In comparison to recent data, both models
produce similar discrepancies. In the light nuclei region, both models
reproduce the data within a factor of 2, for a wide selection of
nuclei with atomic numbers ranging from $Z = 36$ to $Z = 43$. In the
region of very heavy nuclei, the FRDM overestimates the half-lives of
nuclei below the $Z =82$ closed proton shell, while the present work
provides a reasonable description of the decay properties. However,
around the $N = 82$ shell, the D3C$^{*}$ model significantly
underestimates the half-lives. Extending the analysis to the whole
neutron-rich part of the nuclear chart, we obtain very good agreement
with the data, except in the regions just above a closed proton shell,
but below the next closed neutron shell, i.e. the same regions where
the model predicts high contribution of the first-forbidden
transitions. We also performed a statistical analysis, where the
studied the quality of the half-life predictions in contrast to the
measurements, and depending on the type of nuclei. We obtain a very
good average reproduction of the data, with a tendency to overestimate
the half-lives of odd-Z nuclei. In contrast with this result, the very
good overall agreement of the FRDM predictions with the measurements
is a consequence of a cancelation of a significant overestimate of
the half-lives of even-Z (especially even-even) nuclei with a smaller
underestimate of the half-lives of odd-Z nuclei. In both cases, the
description improves for shorter lived nuclei, due to the importance
of the lepton phase space that increases sharply with larger
$Q$-values.

The consistent inclusion of the first-forbidden transitions in this
study has allowed for a systematic study of the contribution of parity
changing transitions to the decay rate for a large number of
nuclei. In specific cases, namely the $N = 50$, $N = 82$ and $N = 126$
isotonic chains, the results of the present study agree well with date
and results based on other models, except in particular cases such as
the nuclei above the $Z = 28$ proton shell where we predict a strong
contribution of the forbidden transitions. For $N = 126$ isotones, the
model predicts an increase of the contribution of first forbidden
transitions to the total decay rate, in agreement with the results
obtained by other models. For nuclei with $N<126$, we predict a smooth
increase of the contribution of the parity changing transitions from
the valley of stability towards the neutron drip-line, and a
saturation at approximately 40\%-60\% of the total rate. Nuclei
with $N>126$ show a suppression of Gamow-Teller transitions that
became only possible by isospin breaking terms mainly due to coulomb. 

Closely related to $\beta$-decay is the process of $\beta$-delayed
neutron emission, which also affects the late stages of the r-process
by contributing to the neutron flux and the redistribution of
mass. The characteristic quantity is the average number of emitted
neutrons which is a measure of the decay rate that falls above the
neutron separation energy. This quantity depends on both the $Q$-value
and the one, two and more neutron separation energies and displays
odd-even staggering both between the isotopic chains and along a
particular chain. We have extracted the probabilities for emission of
one or more neutrons and the average number of emitted neutrons from
the previously calculated decay data. The results show a physical
saturation of the average number of emitted neutrons closer to the
neutron drip-line which indicates that the decaying matter could
provide a significant number of neutrons. In contrast, the FRDM
predicts a similar but overall lower number of emitted neutrons per
decay, with oscillating behaviour close to the neutron drip line.

The present results have been applied in an r-process calculation in a
hot and cold r-process scenarios corresponding to matter ejected under
moderately high entropy. We find substantial changes in the r-process
abundances that are mainly due to the fact that for r-process nuclei
the present half-lives are systematically shorter than those computed
by the FRDM+QRPA approach. 

This study has shown that the theoretical framework based on the
relativistic nuclear energy density functional is a mature one,
capable of providing a good description of sensitive physical
quantities. However, the description of half-lives of a large range of
nuclei remains challenging, especially for nuclei close to the valley
of stability. Further advances in this field will require effort along
three possible directions: (i) improving the relativistic energy density functional will serve to provide
more accurate $Q_{\beta}$ values and enhance the description of the
energy generation during the decay released in the form of electron,
antineutrinos and gamma-rays; (ii) a consistent treatment of
deformations will be necessary to enhance the description of the
transitions in particular open-shell nuclei; and (iii) inclusion of
complex-configurations in the QRPA basis will enrich the transition
spectrum, especially so in the low-energy region, and provide a better
description of beta-decays, but also give information on the detailed
structure of the transition strength.

\begin{acknowledgments}
  This work was supported in part by the Helmholtz International
  Center for FAIR within the framework of the LOEWE program launched
  by the State of Hesse, by the BMBF-Verbundforschungsprojekt number
  06DA7047I, the Helmholtz Association through the Nuclear
  Astrophysics Virtual Institute (VH-VI-417), by the Ministry of
  Science, Education and Sport (MZOS) Project No. 1191005-1010, by
  the IAEA Research Contract No. 18094/R0, and by the DAAD-MZOS bilateral collaboration program. 
\end{acknowledgments}

\bibliography{biblio}

\end{document}